\documentclass[final,3p]{elsarticle}
\usepackage{natbib}
\usepackage[utf8]{inputenc}
\usepackage{url}

\usepackage{import}
\usepackage{tabularx}
\usepackage{comment}
\usepackage{listings}
\usepackage{bm}
\usepackage{caption}
\usepackage{subfigure}
\usepackage{fancyhdr}

\usepackage[framemethod=TikZ]{mdframed}
\usepackage{footnote}
\usepackage{hyperref}

\mdfdefinestyle{style1}{
innerleftmargin=0.5cm,innerrightmargin=0.4cm,
innertopmargin=0.3cm ,innerbottommargin=0.3cm,
roundcorner=10pt,linewidth=1.0pt,
footnoteinside=false}

\begin{document}

\begin{frontmatter}

\title{Evolution of statistical analysis \\in empirical software engineering research: Current state and steps forward}

\author[cth]{Francisco Gomes de Oliveira Neto\corref{cor1}}
\ead{gomesf@chalmers.se}

\author[cth]{Richard Torkar}

\author[cth,bth]{Robert Feldt}

\author[cth]{Lucas Gren}

\author[usi]{Carlo A.\ Furia}

\author[cth]{Ziwei Huang}

\address[cth]{Chalmers and the University of Gothenburg, SE-412 96 Gothenburg, Sweden}
\address[bth]{Blekinge Institute of Technology, SE-371 79 Karlskrona, Sweden}
\address[usi]{Universit\`a della Svizzera italiana, CH-6900 Lugano, Switzerland}
\cortext[cor1]{Corresponding author}

\begin{abstract}
Software engineering research is evolving and papers are increasingly based on empirical data from a multitude of sources, using statistical tests to determine if and to what degree empirical evidence supports their hypotheses. To investigate the practices and trends of statistical analysis in empirical software engineering (ESE), this paper presents a review of a large pool of papers from top-ranked software engineering journals. First, we manually reviewed 161 papers and in the second phase of our method, we conducted a more extensive semi-automatic classification of papers spanning the years 2001--2015 and 5,196 papers.

Results from both review steps was used to: i) identify and analyse the predominant practices in ESE (e.g., using $t$-test or ANOVA), as well as relevant trends in usage of specific statistical methods (e.g., nonparametric tests and effect size measures) and, ii) develop a conceptual model for a statistical analysis workflow with suggestions on how to apply different statistical methods as well as guidelines to avoid pitfalls.

Lastly, we confirm existing claims that current ESE practices lack a standard to report practical significance of results. We illustrate how practical significance can be discussed in terms of both the statistical analysis and in the practitioner's context.
\end{abstract}

\begin{keyword}
\texttt{empirical software engineering, statistical methods, practical significance, semi-automated literature review}
\end{keyword}

\end{frontmatter}

\section{Introduction}\label{sec:intro}
Empirical software engineering (ESE) researchers use a variety of approaches such as statistical methods, grounded-theory, surveys, interviews when conducting empirical studies.\footnote{Even though the approaches mentioned cover Software Engineering (SE) as a whole, this paper focuses on the broad branch of all SE that is mainly interested in empirical data, i.e. Empirical Software Engineering (ESE).} Statistical methods, particularly, are used for interpretation, analysis, organization, and presentation of data. At a minimum, researchers would like to ascertain that their findings are statistically significant\footnote{Even though the common approach of judging statistical significance by the use of $p$-values have recently come under severe scrutiny~\cite{wasserstein2016asa} we here talk about statistical significance in a broader sense of the word.}. Their ultimate goal can be, however, assessing \textit{practical significance}, i.e., that their findings are not only sound but have a non-negligible impact on software engineering practice, thus connecting effort and ROI to the findings.\footnote{In literature, Return-On-Investment refers to, in various ways, the calculation one does to see the benefit (\emph{return}) an investment (\emph{cost}) has.} Therefore, a practitioner's perspective is decisive when defining and analysing an empirical study. Otherwise, reporting statistically significant $p$-values alone does not necessarily imply that the findings are relevant in practice~\cite{KitchenhamPPJHER2002guidelines}. 

To this end, we are in need of robust statistical methods \citep{Kitchenham2017robust} to say, as accurately as possible, the most from the collected data. Nonetheless, creating a connection between the quantitative and qualitative results achieved and the practical implications conveyed by actual software engineering practitioners compel researchers to carefully connect all the elements in their empirical software engineering toolbox (collected data, statistical tests, prior information about the context, etc.). For instance, there are numerous ways to discuss practical significance in terms of statistical methods, such as using coefficients in a multiple linear regression model or using effect size measures (e.g., Cohen's $d$). 

In our experience, 
however, practical significance is rarely explicitly discussed in ESE research papers. Some combination of characteristics that are typical of ESE data complicate the analysis of practical significance: small sample sizes, disparate types of data, which are hard to analyze in a unified framework, and limited availability of general data about software engineering practice.

However, just as ESE is evolving as a research field, so is statistics as a whole. The ever increasing availability of computing power has allowed statistics to branch out into a new sub-discipline, i.e., computational statistics. This has made it possible for researchers in other fields to move away from simple linear models to adopt new and more sophisticated statistical models such as generalized linear models and multilevel models. Increased computational power has also made resampling techniques such as Gibb's sampling~\citep{GemanG84Gibbs}, accessible. In short, just as statistics is evolving, so should ESE research and its use of statistical methods. We thus need to more clearly understand what our current standards for statistical analysis are and in what ways they could evolve.\\

\noindent{\textbf{Problem and proposed solution:}}

Certainly, knowing \textit{which} statistical techniques are applicable is not enough. More importantly, researchers must also know why a particular technique helps to clarify practical significance of results. Using robust statistical methods in ESE research increases the possibility to validate relevant findings and decreases the risk of missing relevant conclusions or drawing the wrong conclusions. This robustness is, currently, strongly connected to the traditional view of Type I and Type II errors, which has led some research fields to scrutinize not only the use of parametric statistics, as has the ESE community~\citep{ArcuriB2011hitchhiker,Kitchenham2017robust}, but also frequentist statistics as a whole. Issues such as the arbitrary $\alpha=.05$ cut-off, the usage of $p$-values and null hypothesis significance testing (NHST), and the reliance on confidence intervals have all been criticized~\citep{Krishna2018smells_analytics,Ioannidis05false,MoreyHRLW2016CI,Nuzzo14errors,TrafimowM15editorial,Woolston15pvalues,wasserstein2016asa}. When analyzing the arguments, we find that many of the issues related to statistical analysis found in other fields (e.g., psychology, biology, and medicine) are equally relevant to ESE research.

In order to address the problems above, the purpose of this paper is to: $i$) provide a view of the evolution of statistical analysis within the ESE literature, and $ii$) to point out aspects that can be improved when applying different statistical methods in ESE research. We focus our discussion on \textit{frequentist} statistical techniques, since they remain prevalent in ESE research. However, one of our conclusions is that long-term improvement in statistical analysis requires adoption of other statistical techniques, so that researchers expand their analysis toolkit when dealing with different validity threats (e.g., missing data points, or heterogeneous data). In fact, Furia et al.\ contrast frequentist and Bayesian data analysis recommending the latter given its  potential to step up generality and predictive capabilities while, in some cases, even providing different outcomes in the analysis~\cite{Furia2018bayesian}. \\

\noindent{\textbf{Method and contributions:}}

Our method includes two literature review phases. First, we present results from a manual review of 161 papers in five popular and top-ranked software engineering journals for the year 2015. The purpose of the manual review was two-fold: $i$) To provide an in-depth view of the current state of the art concerning: the usage of statistics and the frequency and depth of arguments concerning practical significance; and $ii$) act as ground truth to define which statistical methods we could use for the second review phase.

During the second phase, we perform a semi-automatic analysis of a much larger pool of 5,196 papers from the same journals, where we discuss, within our sample, what are the different quantitative approaches used and trends observed across 15 years of ESE research. For instance, after 2011 we identified an increase in the number of papers reporting tests for normality (e.g., Kolmogorov-Smirnov, Shapiro-Wilk) and nonparametric test (e.g., Mann-Whitney's $U$ and Kruskal-Wallis). 

The results are used to build a conceptual model for ESE research where we discuss how different statistical methods can be combined to help researchers to more systematically argue for practical significance. The ESE literature already provides a variety of guidelines on how to apply different statistical approaches. Particularly, our contributions to ESE literature are:
\begin{itemize}
    \item A manual review of 161 papers focused on quantitative research where we extract how different researchers use and describe statistical methods and the corresponding results achieved.
    \item A tool named \texttt{sept} able to semi-automatically analyse papers and extract usage of statistical methods if expressed in common language typically used by researchers. Our results indicate that the semi-automatic extraction can be a viable alternative to a systematic mapping.
    \item An analysis of 15 years (2001--2015) of ESE papers from five selected journals where we identify trends and the main statistical methods used by ESE researchers.
    \item A conceptual model for statistical analysis workflows that indicates what are the different statistical methods that ESE researchers can choose from, as well as the respective guidelines to avoid pitfalls and achieve more clear analyses of practical significance. Moreover, we also include a list of suggested statistical methods that can be an alternative to current practices focusing frequentist analysis.
\end{itemize}

This paper is structured as follows: In Section~\ref{sec:rw} we discuss related work and how our work builds on it, followed by our methodology (Section~\ref{sec:manual}) and the details of our tool and process for semi-automated analysis of papers (Section~\ref{sec:semi-auto}). The results of our entire review process are presented in Section~\ref{sec:results} and then organized into our conceptual model (Section~\ref{sec:ssm}). In Section~\ref{sec:threats} we answer our research questions in terms of our contributions above and discuss threats to validity. Finally, we summarize our conclusions and discuss future work in Section~\ref{sec:concl}.


\section{Related work}\label{sec:rw}
Related work for this study can be divided mainly into two parts: $i$) Guidelines\slash frameworks for reporting studies in software engineering, including studies presenting statistical methods for software engineering; and $ii$) semi-automatic extraction in systematic literature reviews (SLRs).

Over the years, several books and studies have guided ESE researchers on how to design and execute experiments~\citep{WohlinRHOR2012exp} and other types of studies, such as case studies~\citep{RunesonH2008CS} and grounded theory~\citep{Stol2016_grounded_theory}. Some studies focus on certain aspects of the experimental process, e.g., subject experience~\citep{Tantithamthavorn2018icse-seip,HostWT2005expcontext}, bias~\citep{ShepperdBH14bias}, replication~\citep{JuristoV2009replication}, or reporting~\citep{Krishna2018smells_analytics,Jedlitschka2008exp,JedlitschkaJR2014prof}. The early work on guidelines on experimentation by \citet{KitchenhamPPJHER2002guidelines}, and the corresponding work on how to evaluate them~\citep{Kitchenham2006evalguide}, is especially relevant as researchers refine those guidelines as new studies are performed~\cite{Wang2016_icse,Kitchenham2019practices,Tantithamthavorn2018icse-seip}.

All such studies point to different efforts on how to systematically, transparently and repeatably design and report on empirical studies in software engineering. However, one aspect that is often only summarily discussed is what type of statistical methods one should employ and what practical significance is (for instance, \citet{KitchenhamPPJHER2002guidelines} points out that researchers always should try to distinguish between practical and statistical significance). In their study, \citet{Tantithamthavorn2018icse-seip} identify a set of challenges related to practical significance. Particularly, researchers should manage the expectation of practitioners in how they can adopt the published approaches to their own context. Authors also report on pitfalls related to using certain statistical methods such as ANOVA or neglecting to include control variables when designing experiments.

Pitfalls when using statistical methods have also been reported by \citet{ArcuriB2011hitchhiker} where the usage of parametric statistics,\footnote{Where sampling is done from a population following a known probability distribution, e.g., the Gaussian distribution.} especially when it comes to conducting experiments with stochastic elements, should be discouraged in favor of nonparametric statistics.\footnote{Which makes no assumptions about any probability distribution.} A recent study by~\citet{Kitchenham2017robust} followed along that line of thought and suggested statistical methods for ESE research that are more robust. However, existing methods in ESE are still based, predominately, on frequentist analysis, whereas there is little mention of Bayesian data analysis. In fact, over-reliance on null hypothesis significance testing can stray the focus away from obtaining the actual magnitude of a statistical effect~\cite{Krishna2018smells_analytics}, leading to selective reporting of results~\cite{Jorgensen2016133improve_research}.

In the literature there is hardly any explicit discussion of how to introduce Bayesian data analysis in the design, execution, and reporting of software engineering experiments. As far as using Bayesian data analysis as an alternative to frequentist approaches, \citet{Neil1996} published a study already in \citeyear{Neil1996}, which later was followed by more studies. The mentioned studies all apply Bayesian data analysis, but they provide minimal practical guidelines on how researchers should go about in actually \emph{using} Bayesian data analysis. Only recently, \citet{Furia2018bayesian} propose such guidelines after reanalysing two empirical studies with Bayesian techniques revealing its advantages in providing clearer results that are simultaneously robust and nuanced. In turn, \citet{ernst18MLM} presents a conceptual replication of an existing study, arguing that Bayesian multilevel models support cross-project comparisons while preserving local context (mainly through the concept of partial pooling).

To conclude the first part of related work: The introduction of more rigorous statistical methods is wanted, but they should be introduced in the context of existing guidelines\slash frameworks. \citet{Krishna2018smells_analytics} discuss such guidelines focusing on reporting practices and identifying ``bad smells'' in empirical studies within software analytics. Authors argue that such ``bad smells'' can be easily avoided by raising awareness of both best and worst practices in conducting empirical studies. To this end, we discuss our conceptual model in terms of practical advice on how to choose and apply different statistical methods (mainly frequentist approaches), as well as a small set of complementary techniques (e.g., Bayesian data analysis, imputation and causality analysis).

For the second part of the related work we focus on semi-automatic extraction in systematic literature reviews (SLRs) or systematic mappings. Already in 2006 \citet{CohenHPY2006extraction} made the first attempts in applying text mining technologies to reduce the screening burden during reviews; more recently, a systematic review~\citep{Mara-Eves2015extraction} analyzing text mining in systematic reviews indicated that the application of text mining technologies can reduce the workload of systematic reviews by 30--70\% (while recall falls to 95\%). \citet{Mara-Eves2015extraction} also point out that ``it is difficult to establish any overall conclusions about best approaches.'' 
Early work on applying text mining technologies to SLRs in software engineering exists~\citep{Octaviano2015semi-auto,YuKM2016readingSLRs}, but it is still unclear whether this is at the level of maturity required to be a `best' practice.

While it is currently possible to extract primary studies using semi-automatic approaches, the actual synthesis (which an SLR should consist of) still requires substantial human intervention~\citep{Tsafnat2014}.\footnote{Primary studies are the objects of study in SLRs that are later used for extraction of data\slash results\slash conclusions, which in its turn is used for synthesis.} The semi-automatic extraction of primary studies can still help a lot; in particular, it could be used by systematic mapping studies, which include no synthesis. In the next sections, we present one possible way to do this.

\section{Methodology}\label{sec:manual}
\begin{figure}
    \centering
    \includegraphics[width=0.9\textwidth]{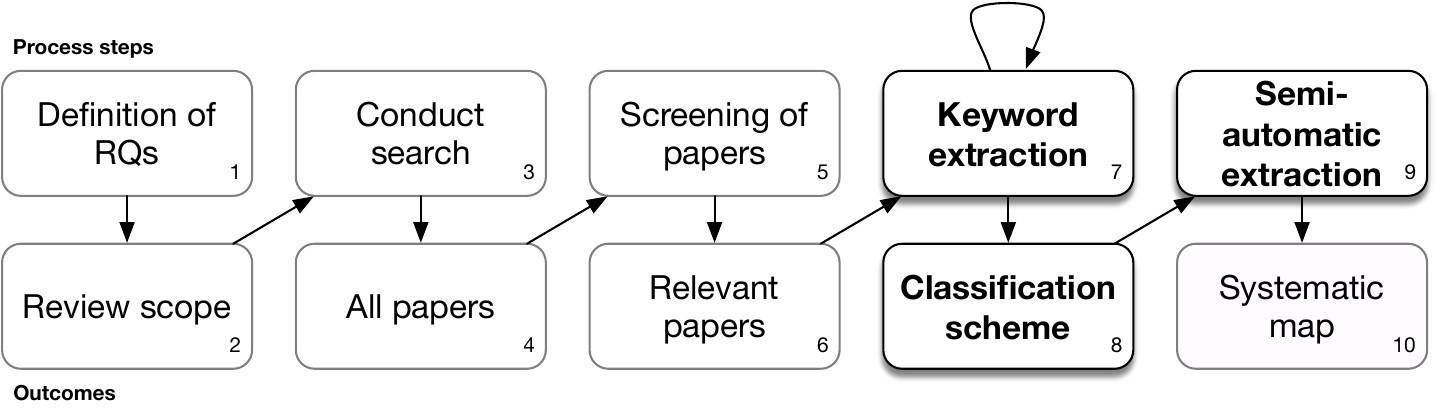}
    \caption{The extraction process, adapted from~\citet{PetersenFMM2008sysmap}, with the parts in bold indicating activities and output that significantly differ from~\cite{PetersenFMM2008sysmap}.}\label{fig:method}
\end{figure}

The current section describes our overall methodology (Figure ~\ref{fig:method}) which extends \citet{PetersenFMM2008sysmap}'s with a semi-automatic part. Additionally, we discuss the scope and details of our manual review process (steps 1--6), the validity of such process (step 7--8) and the overall process for our semi-automated extraction (step 9). Our findings (step 10) are discussed in the upcoming sections of the paper.

\subsection{Scoping and Research Questions}

The overall goal of our literature analysis is to systematically map a large number of journal papers to identify common practices and trends in statistics in ESE research, in particular to address questions of significance. Our investigation relies on two main constructs that support our methodology: $i$) statistical methods and $ii$) empirical papers. 

There are various methods to analyse both quantitative and qualitative data in empirical studies, yet our focus is on quantitative methods. Concerning the definition of empirical, we argue that position papers, editorials or purely theoretical works based on \textit{conceptual} justifications should not be classified as empirical. In contrast, an evaluation\slash validation of a process, technique, idea, can generally be considered as empirical. In practice, one could argue that there are different degrees of empiricism; a view we sympathize with.

Ultimately, empirical studies entail the collection and analysis of data from an observable phenomena~\cite{WohlinRHOR2012exp}. In software engineering, this implies performing experiments, case studies, and surveys on, e.g., projects in industry, open source repositories~\citep{Jaccheri2007opensource_ese} or with students~\citep{Falessi2018students_ese,Feldt2018emse_students}. In short, we propose the the following definitions for statistical methods and empirical papers that set the scope of our research:

\begin{mdframed}[style=style1]
\noindent \textbf{Statistical methods}, or simply \textbf{statistics}, are the models, mathematical formulas and techniques used in the analysis of numerical raw research data.\\

\noindent An \textbf{empirical paper} that makes use of, or potentially \textit{could} make use of, statistical methods as a tool to support its arguments is an empirical study in our context.
\end{mdframed}

In other words, we include studies that report statistical tests (e.g., $t$-test, Mann-Whitney $U$ test), effect size and other quantitative approaches on observable data.\footnote{The list of all methods we consider in this paper is described in Section~\ref{sec:manual}.} Even though our methodology does not cater for qualitative methods (e.g., thematic analysis, grounded theory), the definitions above are inclusive for both quantitative and qualitative studies.\footnote{For instance, when researchers measure inter-rater agreement or perform statistical tests from interview data} Even though descriptive statistics, tables, and plots are relevant for statistical analysis of data, we do not count them in our review process in order to focus our discussion on more complex statistical methods. 

In order to establish a connection between the empirical guidelines and the statistical methods, we also review and discuss the practices of reproducibility and discussions of practical significance. Particularly, we answer the following research questions:

\begin{mdframed}[style=style1]
\begin{itemize}
    \item[\textbf{RQ1:}] What are the main statistical methods used in ESE research, as indicated by studies published in main journals?
    \item[\textbf{RQ2:}] To what extent can we automatically extract usage of statistical methods from ESE literature?
    \item[\textbf{RQ3:}] Are there any trends in the usage of statistical techniques in ESE?
    \item[\textbf{RQ4:}] How often do researchers use statistical results (such as statistical significance) to analyze practical significance?
\end{itemize}
\end{mdframed}

\subsection{Screening and selection of papers}

For this paper, we do not include conference or workshop papers in our review process. We believe that the limited number of pages available to papers in conferences could hinder researchers to report thoroughly on their empirical studies, e.g., not including enough details on the choice and usage of statistics. Moreover, the sample would become heterogeneous, hence introducing many construct and conclusion validity threats to our review process. Since we do not aim for a complete systematic literature review, we make no claims that the sample we use is complete.

Given our focus on journals, we extracted data from: \textit{Transactions on Software Engineering} (TSE), \textit{Transactions on Software Engineering and Methodology} (TOSEM), \textit{Empirical Software Engineering} (EMSE), \textit{Journal of Systems and Software} (JSS), and \textit{Information and Software Technology} (IST). The same sample of journals was used in previous studies by \citet{Kitchenham2019practices}. The main reason for selecting these five journals is that they are well-known and top-ranked software engineering journals focusing primarily on applied scientific contributions. They are also general, in the sense that they do not focus on some particular sub-field; selecting them should thus give a broader overview of the area as a whole. Furthermore, their impact factors and number of citations indicate that these journals publish generally high-quality articles. Even though impact factors and citations are not the only way to assess quality, they remain at least partially reliable indicators of quality. Also based on our knowledge of the journals, we believe they adequately represent the state of the art in ESE research.

Searching the five journals for all publications in 2015 we retrieved a total of 480 publications. Each paper was then screened by title and abstract individually by a researcher (four authors was involved in this step, in total) in order to remove editorials, position papers and secondary studies. Then, among the remaining papers, we performed a round of detailed reviews where researchers read the full paper to remove non-empirical papers, resulting in our set of 313 papers: JSS (101), IST (101), TSE (50), EMSE (41), and TOSEM (20). Details are presented in Table~\ref{tab:screening}. Any uncertainties were discussed among all authors in a workshop to make sure that we did not discard studies that could be seen as empirical.

\begin{table}
    \centering
    \caption{Summary of the paper screening. The rightmost column includes the percentage of empirical papers using statistics for our 2015 sample.}
    \label{tab:screening}
    \begin{tabularx}{0.8\textwidth}{r|rrr|r}
    \hline
                & \textbf{All papers} & \textbf{In scope} & \textbf{Empirical} &  \textbf{Empirical w/ statistics}\\ 
\hline
\textbf{JSS} & 180 & 168 &   101 &   50 (49.5\%) \\ 
 \textbf{IST} & 158 & 129 &  101 &   46 (45.5\%) \\ 
 \textbf{TSE} & 63 & 59 &    50 &   23 (46.0\%) \\ 
 \textbf{EMSE} & 55 & 47&    41 &   30 (73.2\%) \\ 
 \textbf{TOSEM} & 24 & 22&   20 &   12 (60.0\%) \\ 
 \hline
 $\sum$ & 480 & 425 & 313 & 161 (51.4\%) \\ 
 \hline
\end{tabularx}
\end{table}

\subsection{Validity of our review processes}

To later reproduce this study one can find the processed data of our reviews online\footnote{\url{https://goo.gl/AZnqMJ}}. We next analyzed the papers according to the degree of empiricism, to get an estimate on the usage of statistics. More than a quarter of the papers in scope (112 out of 425, i.e., 26.3\%) were classified as non-empirical and roughly half of the remaining, empirircal papers (161 out of 313) actually use statistics\footnote{Most qualitative studies were classified as empirical, but the degree of usage concerning statistics varied considerably.} In the remainder of this paper, we call \textit{primary studies} those that are empirical \textit{and} use statistics. This means 161 papers or 37.9\% of the manually screened papers.

In our systematic mapping process, we wanted to identify keywords from the papers that indicate whether certain statistical methods (such as parametric tests, effect sizes, and so on) were used. To this end, we built a questionnaire to guide the manual extraction.\footnote{The questionnaire is available online \url{https://bit.ly/2IxadBU}} 

During several iterations the questionnaire was refined, used and evaluated by four of the authors. The reliability of the questionnaire was analyzed quantitatively through an inter-rate reliability measure on the answers. These measures are used to verify the consensus in the ratings given by various raters. Different measures have different thresholds in literature that represent the level of agreement between raters. We use Krippendorff's $\alpha_K$ due to its following properties~\citep{hayesK07reliability}:

\begin{itemize}
    \item Assesses the \textit{agreement} between $\geq2$ independent reviewers who conduct an independent analysis.
    \item Uses the distribution of the categories or scale points as
used by the reviewers.
    \item Consists of a numerical scale between two points allowing a sensible reliability interpretation.
    \item Is appropriate to the level of measurement of the data.
    \item Has known, or computable, sampling behavior.
\end{itemize}

Two alternatives to Krippendorff's $\alpha_K$ are Cohen's $\kappa$~\citep{Cohen60kappa} or Cronbach's $\alpha_C$~\citep{Cronbach1951alpha,hayesK07reliability}. However, in our study, the two latter measures violate the first item (reviewers are not freely exchangeable and permutable) and third item (setting a zero point as in correlation\slash association statistics).

The validation of the questionnaire made use of a not fully crossed design~\citep{PerronG2015measurement}. A random sample of 18 papers (\textgreater 11\% of the 161 primary studies) was used to calculate $\alpha_K$. Reviewing 18 papers by four of the authors each gave, $\alpha_K=0.7$, 95\% CI [0.46, 0.90].\footnote{Krippendorff's $\alpha_K$ was calculated using \textsf{R}~\citep{r-ref} with package  \textsf{irr}~\citep{irr-ref}; to control for instrument reliability we also used \href{https://www.ibm.com/analytics/us/en/technology/spss/}{\textsf{SPSS 24}} and the macro \href{http://www.afhayes.com/public/kalpha.zip}{\textsf{KALPHA}}. No significant differences were noticed in the outputs.} For the confidence intervals, we conducted a bootstrap sample (10,000) of the distribution of $\alpha_K$ from the given reliability data. Even though non fully crossed designs underestimate the true reliability~\citep{Hallgren2012IRR-cross,Putka08measurement}, 
and our $\alpha_K$ was above the recommended threshold of 0.667~\citep{krippendorff2004content},
we remained unsure about the accuracy of our estimate due to the wide confidence interval. Therefore, to refine the reliability analysis, we looked at the agreement between different subcategories of statistical methods used in the questionnaire.

We conservatively classified a subcategory to have high reliability if two reviewers independently of each other selected precisely the same check-boxes in a subcategory when reviewing the same paper. As an example, if a paper had a test for normality (i.e., distribution tests), the raters should have reported precisely the same distribution tests in the questionnaire. This way, we calculated the reliability of each subcategory individually, i.e., we would get a number on how much trust we could put into using those keywords in our semi-automatic review.

We see in Table~\ref{tbl:alpha} that some subcategories had $\alpha_K<0.667$, indicating that they may decrease the accuracy if used for the semi-automatic extraction. Therefore, we decided to only use, in the semi-automatic extraction subcategories, $\alpha_K \geq 0.667$ (marked in bold in Table~\ref{tbl:alpha}) or subcategories having strong inter-reviewer agreement but for which we could not reliable calculate $\alpha_K$ due to some statistical anomalies (marked with $\dagger$ in Table~\ref{tbl:alpha}).

\begin{table}
 \caption{Overview of $\alpha_K$ per subcategory (ascending order). Bold text indicates subcategories suitable for semi-automatic extraction, i.e., $\alpha_K \geq 0.667$. An $\alpha_K \geq 0.667$ indicates the lowest conceivable limit where tentative conclusions are still attainable~\cite{krippendorff2004content}.} 
    \label{tbl:alpha}
    \centering
    \begin{tabularx}{\textwidth}{Xccc}
    \hline 
    \textbf{Subcategory} & $\bm{\alpha_K}$ & \textbf{with 95\% CI} & \textbf{ratio} \\
    \hline 
         Repeatability                      & 0.22 & [-0.01, 0.45]  & ---\\
         Practical significance             & 0.29 & [0.08, 0.50]   & ---\\
         Raw data availability              & 0.47 & [0.23, 0.68]   & ---\\
         Type I correction                  & 0.61 & [0.36, 0.83] &---\\
         \hline
         \hline
         \textbf{Power analysis} & $\bm{0.70}$ & $\bm{[0.47, 0.90]}$ &---\\
         \textbf{Effect sizes} & $\bm{0.74}$ & $\bm{[0.61, 0.87]}$ &---\\
         \textbf{Distribution tests} & $\bm{0.87}$ & $\bm{[0.70, 1.00]}$ &---\\
         \textbf{Nonparametric tests} & $\bm{0.92}$ & $\bm{[0.83, 1.00]}$ &---\\
         \textbf{Latent variable analysis}$\dagger$ & n\slash a & n\slash a& $\bm{14/18}$\\
         \textbf{Quantitative analysis}$\dagger$ & n\slash a&n\slash a & $\bm{17/18}$\\
         \textbf{Parametric tests} & $\bm{1.00}$ & n\slash a & ---\\
         \textbf{Confidence intervals} & $\bm{1.00}$ & n\slash a & ---\\
         \hline
    \end{tabularx}
\end{table}

The reason we have two instances marked with $\dagger$ is due to the very conservative approach $\alpha_K$ uses when calculating inter-rater reliability~\citep{Guangchao15}. When it comes to those two categories, we had an agreement of 78\% and 94\%, respectively, and yet $\alpha_K=0$. We opted for including the two categories in the semi-automatic extraction since the agreement was convincing.

In extreme cases, even if referees have an agreement of \textgreater95\%, it could still lead to a negative $\alpha_K$ if dealing with significantly skewed distributions. In other words, $\alpha_K$ is a chance-corrected measure, and an $\alpha_K$ of zero means that agreement observed is precisely the agreement expected by chance. In our case, there was little variation on how different researchers classified the papers, so the expected agreement of several categories was very high (close to 100\%).

\subsection{Classification scheme and keyword extraction}
After validating the manual review process, we used the text from papers within each category to create a classification scheme used by our semi-automated extraction tool. For each desired property of a paper (e.g., a parametric test), the tool user writes an extractor for such a property in a simple domain-specific language. The extractor is composed of manually marked textual examples from several papers that contain said property.

The tool then uses those examples to systematically extract and score \textit{similar information} (e.g., analysis using $t$-test) from other papers. The tool then aggregates the specific statistical methods into the categories deemed reliable from Table \ref{tbl:alpha}. Note that the extraction and scoring is not based only on plain keywords, instead the tool checks the context where the keyword was found, i.e., the sentence and adjacent paragraphs around the keywords. 

We call the extraction ``semi-automatic'' due to the human involvement in defining the schemes and context for the categories. Still, the scheme created from our 2015 sample was used to extract data from a much larger sample of papers ($N=5,196$), spanning across the years 2001--2015. The design of such schemes, as well as validity of our extraction tool is detailed in the next section.

We use results from both review processes (manual and semi-automatic) to support our claims about the evolution of statistical methods in ESE (Section~\ref{sec:results}) and the design of our conceptual model (Section~\ref{sec:ssm}).

\section{Semi-automatic review}\label{sec:semi-auto}
Since the manual review is time-consuming and does not scale up, we developed a tool, named \textsf{sept}, that classifies papers based on flexible keywords. Applying \textsf{sept} to a large set of papers allows us to address broader research questions more reliably. Such a tool also provides other benefits such as more objective reviews or, through future extensions, extracting details about the use of different statistical techniques. Potentially, that could be useful not only in retrospect, e.g., to study the papers that are already out there, but also for journals and conferences at submission time to give improvement advice to authors, and at review time to help focus the work of reviewers.

It is important to note that our aim is \textit{not} to create a completely automated tool. Such a tool seems unlikely to \textit{accurately} work in general, since research approaches and ways of describing statistical methods and analyses vary way too much. Our goals for tool support are more modest: creating a tool that can find parts of academic software engineering papers that are likely to indicate the presence (or absence) of specific statistical analysis methods. The general approach used for implementing such a tool should also help in extracting other elements and aspects of papers (e.g., discussions about practical significance or the availability of data for replication and reproducibility). Aa a sidenote, the tool has been adapted to initial screening of violations to Double-Blind Reviewing rules and has been used for this purpose in the ICSE 2018, ICST 2018, as well as ICSA 2018 conferences.

Below we detail \textsf{sept}'s design and how we evaluated whether it can achieve a review accuracy that is close enough to human. The tool is available online as a Docker image, whereas the scripts and artefacts are available as a Zenodo package.\footnote{\url{https://hub.docker.com/r/robertfeldt/sept}}\footnote{\url{http://doi.org/10.5281/zenodo.3294508}} However, due to copyright issues the sample of PDF papers we used in our analysis cannot be made publicly available. Instead, we provide processed data in CSV files along with the tool.

\subsection{Overall design of review tool} 
Initially, the automated extraction was intended to support and check our manual review process using simple text matching to review a larger set of papers and mitigate human errors, such as missing important keywords in the text. The process that several of us had implicitly used in the manual review was partly based on searching for specific key terms and then judging if any of the `hits' gave evidence for the use of the analysis in question. The design and further development of our tool grew naturally from this intuition, since initial results were promising.

To classify papers into different categories (e.g., parametric tests, quantitative analysis, nonparametric tests) we would match different terms (e.g., Student's $t$-test, Mann-Whitney $U$ test, Friedman test). The basic observation was that we could take our decision based on \textit{positive evidence}, i.e., whether a paper used certain statistical analysis, in one or a few consecutive paragraphs of text (often even based on a few sentences around a main match). For example, when looking for the use of a nonparametric test, such as the Mann-Whitney $U$ test, several of the researchers searched for `Mann' or `Whitney' and then manually read the surrounding text to confirm that the test was indeed used. Besides, they would, of course, read specific parts such as results and analysis sections which are more likely to contain evidence of analyses used.

Conversely, sometimes the evidence was presented as \textit{negative evidence}. For example, when authors discuss that \textit{instead} of using a parametric statistical test (e.g., $t$-test) they used a nonparametric test. Nonetheless, for the majority of cases, the evidence found around a match was positive, e.g., a match on `Wilcoxon rank-sum test' was almost exclusively when discussing that such a test had been used and what the outcome had been. Another observation was that some of the researchers sometimes had missed specific analyses during the manual review. In particular, this was the case when it was reported in an unusual fashion, or when the names of the analysis method were misspelled or described in less precise terms.

We thus set out to develop \textsf{sept} to a level where we judged it to give reasonable results compared to manual, human reviews. As argued above we do not think such a tool can ever be perfect; that would require an artificial intelligence with a lot of experience from reading and understanding software engineering research and statistical analysis. Instead, we evolved the tool from the exact matching of text to a more fuzzy matching based on examples and heuristics. 

\begin{figure}
    \centering
    \includegraphics[width=0.9\textwidth]{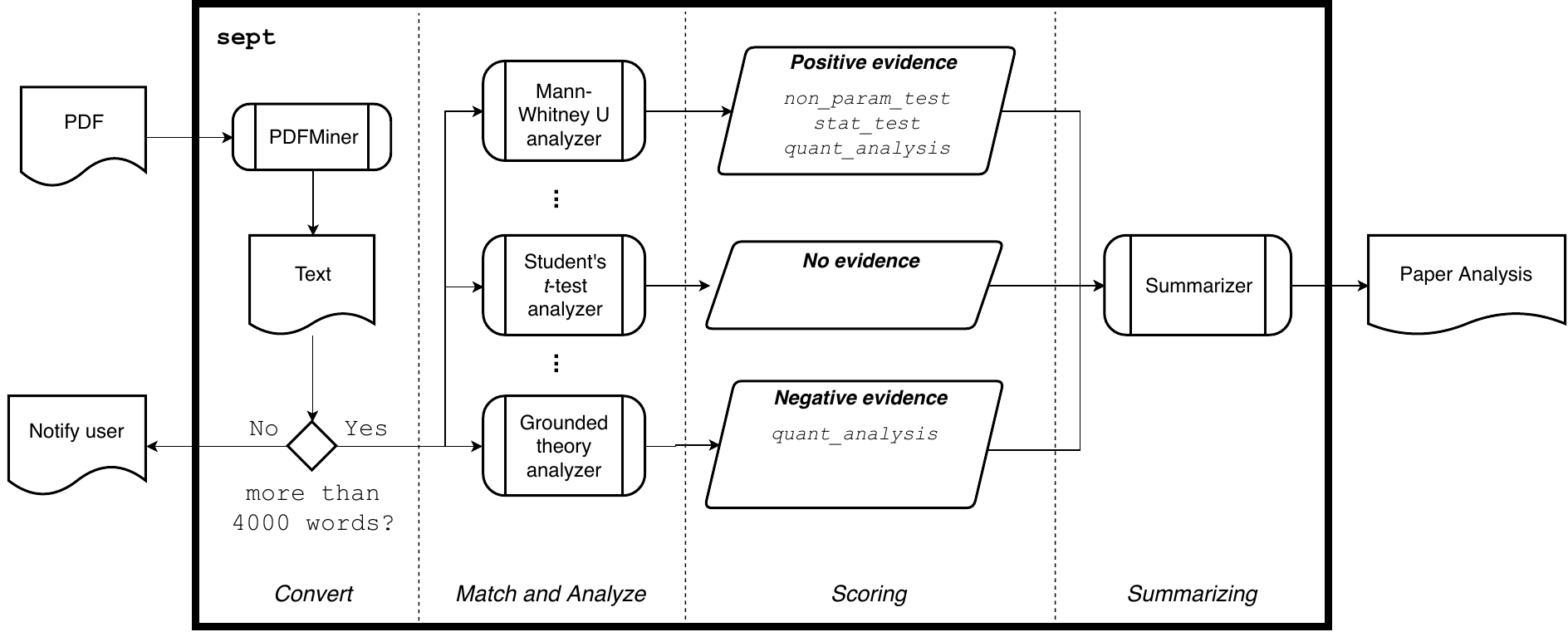}
    \caption{Overview of the \textsf{sept} tool for semi-automated checking of software engineering research papers. The paper (PDF) file is converted to a text file and is then passed to analyzers that do fuzzy matching (three exemplified here). Each analyzer focuses on finding evidence for (\textbf{positive}), against (\textbf{negative}) or finding \textbf{no} evidence (i.e., skip) for the use of a certain type of statistical test or aspect that is used in the paper. The tool outputs an analysis file that details all evidence found and marks what lead the tool to judge the evidence to be present.}
    \label{fig:sept}
\end{figure}

Figure~\ref{fig:sept} shows the overall design of the \textsf{sept} tool. First, we use \textsf{PDFMiner} to extract text from the paper's PDF file.\footnote{\url{https://euske.github.io/pdfminer/}.} In order to discard publications containing only editorials, short or position papers, the tool skips texts with less than 4,000 words and notifies the user.\footnote{The cutoff of 4,000 words was based on an average obtained from few runs with a controlled sample of editorials, short and position papers from the sampled journals in our analysis.} Otherwise, the tool proceeds with matching, scoring and analyzing the distinct terms throughout the text. Finally, it summarizes the analysis results in a report. The \textit{analyzers} and \textit{matchers} describe, respectively, what terms are searched for and how to match them. An example of an analyzer with corresponding matchers is shown in Table \ref{tab:ttestexample}.

\begin{table}
    \centering
    \caption{Example of a text analyzer for the parametric Student's $t$-test in the \textsf{sept} tool. The analyzer is composed of a set of positive and negative examples as well as a skip matcher (here in the form of a regular expression). Additionally, the analyzer includes a list of synonyms for the primary matching terms (enclosed by [[[ and ]]] markers). Matching any of the examples will contribute to scoring of the specified tags.}\label{tab:ttestexample}
\begin{tabularx}{\textwidth}{rXX}
\hline
\multicolumn{2}{c}{\textbf{Student's $t$-test Analyzer}}\\
\hline
\textbf{Positive examples} & 
1. \texttt{We \_\_used\_\_ a [[[Student's t-test]]]} \\
&\shortstack[l]{2. \texttt{the [[[t-test]]] \_\_to test hypotheses\_\_ concerning accuracy; both}\\
\texttt{\_\_with the significance level\_\_$ \alpha$ \_\_set to\_\_ 0.05}} \\
&\shortstack[l]{3. \texttt{The \_\_difference in means\_\_ between pre-3.6 Firefox versions} \\
 \texttt{and Firefox 3.6 versions is about 0.56 watts ([[[T-test]]]} \\
 \texttt{\_\_p-value\_\_ of 0.012).}
 } \\ 
 \textbf{Skip matchers} & \texttt{\#RegexpMatcher(r``[a-zA-Z]\{1\}t(\textbackslash s+|-)test''i)\#} \\
 \textbf{Negative examples} & \texttt{We \_\_did not use\_\_ a [[[Student's t-test]]] to} \\
  \hline
 \textbf{Synonyms} & \texttt{"Student's t test", 
        "Students t test", 
        "Students t-test", 
        "Student t test", 
        "Student t-test", 
        "Student t",
        "Welch's t-test", 
        "Welchs t-test", 
        "Welch's t test", 
        "Welchs t test", 
        "t-test",
        "t test",}
 \\
  \hline
 \textbf{Tags} & \texttt{parametric\_test}, \texttt{statistical\_test}, \texttt{quantitative\_analysis} \\
 \hline
\end{tabularx}
\end{table}

Analyzers use examples taken directly from the papers (e.g., during a manual review), or are written as more general and standard patterns or phrases. Each analyzer has a set of positive examples, negative examples and skip matchers. For each analyzer, the tool first tries to match its positive examples. Then, for all matching targets (i.e., a region of the text), the skip matchers are used to check whether any of the positive matches should instead be skipped, hence refining the matching based on the positive examples. A typical example of skip matching is shown in Table~\ref{tab:ttestexample} for the Student's $t$-test matcher. Several matches become positive for the string `unit test' due to the sub-string `t test' (which is a synonym match defined in Table~\ref{tab:ttestexample}). The skip matching makes sure that this specific match is skipped and not further considered by specifying a regular expression. In practice, the skip matcher is rarely used except for shorter and more general terms that may be matched in the positive example. Note that these matches would be misleading (i.e., false positive) and should not be counted (as discussed in our tool validation subsection).

In turn, the negative examples indicate that a paper does \textbf{not} employ a particular statistical technique. For example,~\citet{garousi2015usage} write in their validity threats section that ``Furthermore, we have not conducted an effect size analysis on the data and results; thus this is another potential threat to the conclusion validity.'' We take this as a strong indication that there is no use of effect size calculations in their analysis. Negative matches thus typically override any earlier positive matches in the final scoring as further described below.

The positive and negative examples are written in a small domain specific language, which allows annotating sentences and text copied directly from a paper. Parts of the sentence enclosed by [[[ and ]]] markers are taken to be the \textit{primary match} target. Similarly, text between \_\_ markers is used to search for \textit{supporting matches} in close vicinity to the primary match. This distinction between primary and supporting matches is important in the next step of the extraction (i.e., scoring matches), since supporting matches increase the score of the matched target and contribute more clearly to the evidence for the analyzer. Supporting matches are used differently in positive and negative examples: when matching a positive example, any number of supporting matches may also match---the more supporting matches that match, the higher the positive evidence; in contrast, when matching a negative example, \textit{all} supporting matches have to match to confirm negative evidence---if this is the case, the negative evidence is then weighted strongly.

After the matching process completes, the analyzers will score each match to estimate how likely it is to provide positive or negative evidence. The outcome of the scoring is a set of tags and the type of evidence found. The tags represent chains of evidence that a certain match and analyzer provides. For example, all the nonparametric statistical tests, such as Mann-Whitney $U$ or Friedman test, provide evidence for the same tags: \verb+non_parametric_test+, \verb+statistical_test+, and \verb+quantitative_analysis+. In other words, positive evidence that a Friedman test has been used gives positive evidence that a statistical test has been used and that quantitative analysis has been carried out in the analyzed paper.

Finally, the \textsf{summarizer} component (Figure~\ref{fig:sept}) collects the evidence and outputs a report that gives a detailed account of the results. This both summarizes which tags we have found evidence of, and presents the textual context around each high-scoring match that supports that evidence. The output is human-readable, so that the results from the tool can be checked by humans.

The design and implementation of our tool are similar to the approach used by~\citet{Octaviano2015semi-auto} for filtering studies to be included (or excluded) in systematic literature reviews or mapping studies. However, they do text matching only on the title, abstract and keywords of a study, while we match throughout the \textit{full text} of each paper. We also found it critical to do fuzzy matching of both the primary match string as well as of parts of the sentence surrounding the main match (specially because it is not uncommon that figure and table text are intermixed in the middle of the text, white-space, and hyphenated words, after converting the PDF of a paper to text). It is not clear if~\citeauthor{Octaviano2015semi-auto} do any fuzzy matching at all. On the other hand, they might not need to, since the meta-data they match on can, at many times, be extracted from databases. 

\subsection{Scoring of evidence in tool}
Our scoring is based on the observation that positive examples are more permissive when matching (i.e., they do not require all supports to match but are rather ranked based on the number of matching supports), whereas once there is any negative evidence, for which all support matches must match, it will outweigh the positive evidence. At the same time, negative examples match rarely because of their more demanding requirements. The final evidence for a given analyzer contributes to all its tags and the final scores per tag are then aggregated.

\begin{table}
 \caption{Example of three analyzers and the total of matching targets. The corresponding matching targets are scored against four different tags: \texttt{quantitative\_analysis}, \texttt{statistical\_test}, \texttt{parametric\_test} and \texttt{non\_parametric\_test}.}
 \label{tab:analyzers}
    \centering
    \begin{tabularx}{\textwidth}{llXcc}
    \hline 
    \textbf{ID} & \textbf{Analyzer Name} & \textbf{Tags} & \textbf{Positive matches} & \textbf{Negative matches} \\ 
    \hline 
    A1 & Mann-Whitney $U$ (MWU)  & \texttt{quantitative\_analysis}, \texttt{statistical\_test}, \texttt{non\_parametric\_test} & 2 & 0\\
    \hline
    A2 & Student's $t$-test & \texttt{quantitative\_analysis}, \texttt{statistical\_test}, \texttt{parametric\_test} & 3 & 1 \\ 
    \hline
    A3 & Cliff's $d$ effect size & \texttt{quantitative\_analysis}  & 4 & 0 \\
    \hline
    \end{tabularx}
\end{table}

For example, consider the three analyzers and their corresponding tags in Table \ref{tab:analyzers}. Each analyzer has some positive and negative matchers that are then aggregated when scoring the tags. All three analyzers contribute to \texttt{quantitative\_analysis}, hence scoring one negative piece of evidence (A2) and two positive pieces of evidence (A1 and A3).\footnote{Note that even though A2 has positive matches for $t$-test, the negative match overrides them.} Meanwhile, the \texttt{statistical\_test} tag has one negative evidence (A2) and one positive evidence (A1). Finally, we will have one negative evidence for \texttt{parametric\_test} (A2) and one positive evidence for \texttt{non\_parametric\_test} (A2). Ultimately, we will then count the paper as doing both quantitative analysis and making use of statistical tests, particularly, a nonparametric test.

While the scoring scheme could be improved, our validation shows a reasonable accuracy. We leave more advanced scoring and experimentation with more sophisticated approaches for future work.


\subsection{Validation of the \textsf{sept} tool}
\label{subsec:tool:validation}

To ensure that the summary information captured by \textsf{sept} has a reasonable accuracy we validated some tags manually in several different rounds. After each round we updated the tool based on false positives and false negatives identified. In our validation, false positives are papers that were classified by the tool as covering a tag (e.g., parametric tests) but do not actually deal with the tag's technique. Conversely, false negatives are papers that include an analysis related to one of the tags, but were not classified as such by the tool. For instance, if a paper has an ANOVA test, but is not matched by the tool we count that ANOVA match as a false negative for the \texttt{parametric\_test} tag. Naturally, we aim to increase the number of true positive and true negatives detected. 

During development and even after reaching a stable version, we validated the analyzers used in the classification. We selected a random subset of reliable categories from our manual review process (Section~\ref{sec:manual}). For reasons of brevity we summarize below all validation rounds and give some details that support our case that the final version of the tool has reached a useful level of accuracy. We focused on validating the classification of the following tags: \texttt{non\_parametric\_test}, \texttt{parametric\_test}, \texttt{multiple\_testing\_correction}, \texttt{normality}, \texttt{data\_available\_online}, \texttt{power\_analysis}. The first four tags were chosen since they could be matched via keywords (e.g., Kolmogorov-Smirnov, $t$-test, Bonferroni), whereas matching the latter two tags is less straightforward where \textsf{sept} needs to analyze the context where the terms are being used. We used stratified random sampling to select papers with the chosen tags. Additionally, we manually added papers where the data would show unusual patterns, i.e., through peaks in specific year intervals (detailed below). Table \ref{tab:valid1} presents the number of papers analyzed for each tag used in the validation.

\begin{table}
 \caption{Distribution of true positive (P), false positive (FP), and true negative (TN) matches before updating the tool based on the additional manual validation. Values in parentheses indicate the results of re-running \texttt{sept} on those papers after refining its analyzers.}
 \label{tab:valid1}
    \centering
    \begin{tabularx}{\textwidth}{X|rr|rr|rr|r}
    \hline
    \textbf{Tags} & \multicolumn{2}{c|}{\textbf{P}} & \multicolumn{2}{c|}{\textbf{FP}} & \multicolumn{2}{c|}{\textbf{TN}} & \multicolumn{1}{c}{\textbf{Total}} \\ 
    \hline
    \texttt{non\_parametric\_test}   & 17 & (18)  & 3  & (2)   & 0  & & 20 \\
\texttt{multiple\_testing\_correction}& 5& (5)  & 1  & (1) & 36 & (36) & 42 \\
    \texttt{normality}               & 26 & (27)& 5  & (4)& 0  & & 31 \\
    \texttt{data\_available\_online} & 21  & (21) & 0  &         & 0  & & 21  \\
    \texttt{power\_analysis}         & 6  & (7)  & 4  & (3)  & 0  & & 10 \\
    \texttt{parametric\_test}        & 1  & (9)   & 9 & (1)  & 0  & & 10 \\
    \hline
    \textbf{Total} (matches)         & 76& (79) & 22 & (11) & 36 & (44) & 134 \\
    \hline
    \end{tabularx}
\end{table}

All 21 matched papers for the online data availability tag (e.g., via URL links or GitHub repositories) were correct. Investigating false negatives was unfeasible,\footnote{An example of a false negative would be authors providing material on their own (or the publisher's) website but not mentioning it in the paper} given the large amount of papers where data \textit{was not} made available. Additionally, we excluded matches where authors would only state that supplementary material was available online without providing links.

The vast majority of the matches for \texttt{normality} and \texttt{non\_parametric\_tests} were true positives. The few false positives for \texttt{non\_parametric\_tests} included two matches where authors state that they could have used the corresponding statistic test (but did not), and one false positive from a systematic review in power analysis. In the case of the normality tags, we found four cases where the test was used to test distributions other than normal (mostly using the Kolmogorov-Smirnov test), and one from generically worded descriptions in the analyzer.

Similar false positives were detected for the \texttt{power\_analysis} tags. For power analysis we identified four false positives, due to: generically worded descriptions in the analyzer (2 cases), power analysis performed only in cited related work (1 case), and claims that power analysis could \textit{not} be performed (1 case). This shows a limitation of our approach, i.e., we do not use advanced natural language processing to try to understand the semantics, but require unique textual elements that we can match on. To mitigate this limitation, we scaled down the power analysis analyzer to target only more specific expressions.

The false positives for \texttt{parametric\_tests} included mainly cases where `$t$-test' and `$F$-test' would match due to substrings in various words (e.g., unit tests, and goodness of fit tests). Those cases were then used to refine the tool, increasing the number of skip matches. In order to check for false negatives, we focused on specific unclear areas of the data where we wanted to be sure the tool was not misleading us. For the \texttt{multiple\_testing\_correction}, we saw an abrupt reduction in the number of papers of this category over the years 2007--2009. We sampled and checked 36 papers from various journals between those years tagged with \texttt{no evidence} of using corrections for multiple testing, and concluded that the matches were indeed correct. We also checked the positive matches and concluded that only one was a false positive where authors state that they did not use the test.

Finally, in the last round, we decided that a number of the remaining problems came from the analysis of systematic reviews where it is often the case that statistical tests and other types of statistical features are indirectly discussed. This is when we decided to add an analyzer to detect systematic reviews and systematic mapping studies. If a paper has positive evidence (and no negative evidence) of being such a secondary study we do not consider it for the final analysis and statistics reported here. In the final run on 5,196 papers, 170 such papers were found. The analyzer for secondary studies is a special one since it considers only text in the first 5\% of the paper text. This is likely to cover the title, abstract and parts of the introduction where it is likely to find enough positive evidence.

We then re-ran the improved tool, validated the same tags again and proceeded to other tags. Furthermore, we added specifically difficult and faulty papers to the tool's test suite to support regression testing in the future evolution of the tool.

\subsection{Limitations of the tool}
The validation described above revealed limitations of the tool's extraction capabilities. For instance, the tool cannot separate the use of a statistical test or algorithm as part of the proposed technique or method proposed by the paper (in the following called the technical level) from the usage of said technique or method (called the methodological level). To evaluate the usage of statistical methods in a paper, we want to focus on the methodological level; whereas the technical level is not our concern.

The positive examples were of course chosen from parts of papers discussing the methodological level, and we did add negative examples for matches on the technical level when we encountered them, but we suspect there are still some such false positives in our data. However, if a paper has the sophistication to employ statistical methods at the technical level it is likely to also use the same or related methods at the methodological level. So, at least for the more general tags, we do not believe this is a major threat to the accuracy of the results.

\section{Results from classification}\label{sec:results}
\begin{figure}
    \centering
    \includegraphics[width=0.8\textwidth]{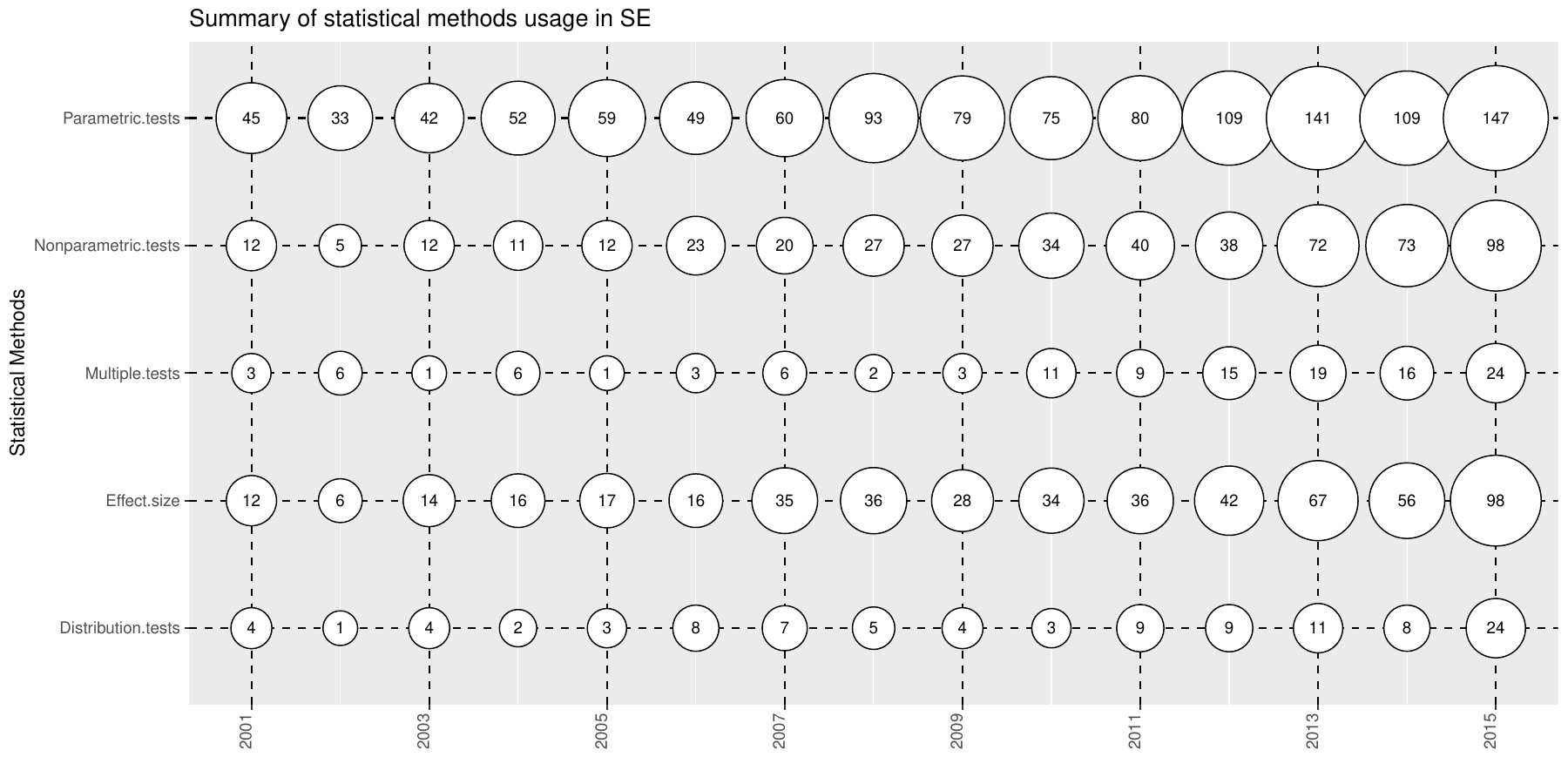}
    \caption{The mapping of different statistical approaches extracted from papers of the selected journals between 2001--2015.}
    \label{fig:sys_map}
\end{figure}

The following sections present results of the manual and semi-automatic classification of studies. The version of our tool used to extract data reported on later in this paper includes a total of 59 analyzers. For space reasons we do not list them all, but they can be found online.\footnote{\url{https://goo.gl/efBdDz}} A summary of the different statistical methods aggregated in categories is presented in Figure~\ref{fig:sys_map}. We motivate the choice of categories in our methodology. Note that the semi-automated extraction is used only in the reliable categories (Table~\ref{tbl:alpha} in Section~\ref{sec:manual}).

The goal is to discuss the main practices in ESE and possible trends in those practices. Our results are visually presented in charts where the $y$-axis is the normalized scores of the number of papers where we found positive evidence,\footnote{The journals differ significantly in terms of the number of published papers per year. Consequently, using absolute numbers would be misleading to our analysis.} i.e., the number of positive evidence divided by the number of papers published that year for each journal. In short, if we would find positive evidence for all papers published in one year, the stacked bar would reach 5.0 (since we have five journals). This allows us to examine an overall trend for all journals by comparing the total height, i.e., the point is to show that a value is the sum of other values, and we are only interested in comparing the totals. Each chart contains also a local regression (we used LOESS, i.e., a locally estimated scatterplot smoothing) line that can indicate trends in the usage of the corresponding statistical method over the years.\footnote{The smoothing of the regression was done using \texttt{geom\_smooth} (ggplot2) in R.}

\subsection{Descriptive statistics}\label{subsec:res:overview}
In the manual classification, 85\% of the studies (classified as empirical and using statistical analysis) were classified as presenting descriptive statistics of some sort, while also reasoning about the data that was plotted or presented in tables. This baseline is encouraging because it shows that most papers have a level of reasoning about their collected data, such as conveying central tendencies (mean, media and mode), dispersion (variance and ranges) and shape (skewness).

\subsection{Power analysis}\label{subsec:res:pwr}
The power of a statistical test can be thought of as the probability of finding an effect that actually exists. Or more correctly, in classical statistics, this means the probability of rejecting the null hypothesis if it is false~\citep{cohen1992statistical}. A statistical power analysis is an investigation of such a probability for a specific study and can be conducted both before (\emph{a priori}) or after the data collection (\emph{post hoc}). An analysis of the \emph{a priori} power is based on previous studies, or the relationship between four quantities: Sample size, effect size, significance level and power. To calculate one of these you need to establish the other three quantities. For \emph{post hoc} analysis, on the other hand, one usually refers to the observed power~\citep{onwuegbuzie2004post}.

\citet{Dyba2006power} reviewed 103 papers on controlled experiments published in nine major software engineering journals and three conference proceedings in the decade 1993--2002. The results showed that \textit{``the statistical power of software engineering experiments falls substantially below accepted norms as well as the levels found in the related field of information systems research.''} In short, to handle Type II errors one should make sure to have a sufficiently large sample size and set up the experiment to see larger effect sizes~\citep{Banerjee09}.

Investigating the output from \textsf{sept} provides us with scarce evidence concerning the usage of power analysis, i.e., the statement by \citet{Dyba2006power} still holds for 2001--2015. We cannot say much about any trends since we have found relatively few cases of power analysis being applied; only seven counts of positive evidence were found, in data spanning 15 years of research over five distinct publication venues. However, in the cases where we found evidence of power analysis being conducted, it invariably was of a \emph{post hoc} nature. At the same time, when validating \textsf{sept} we found that papers with power analysis are harder to match than other tags, and hence this category is more susceptible to false negatives.

\subsection{Distribution tests}\label{subsec:res:normality}
Tests for normality check whether data is likely to come from a normal distribution, but they are part of a larger family of \textit{distribution tests} that can test for different kinds of distributions. A distribution test is typically applied before conducting other statistical tests that only work under certain distributional assumptions (most commonly,  normality~\citep{shapiro1965analysis}).

\begin{figure}
    \centering
    \includegraphics[width=0.4\textwidth]{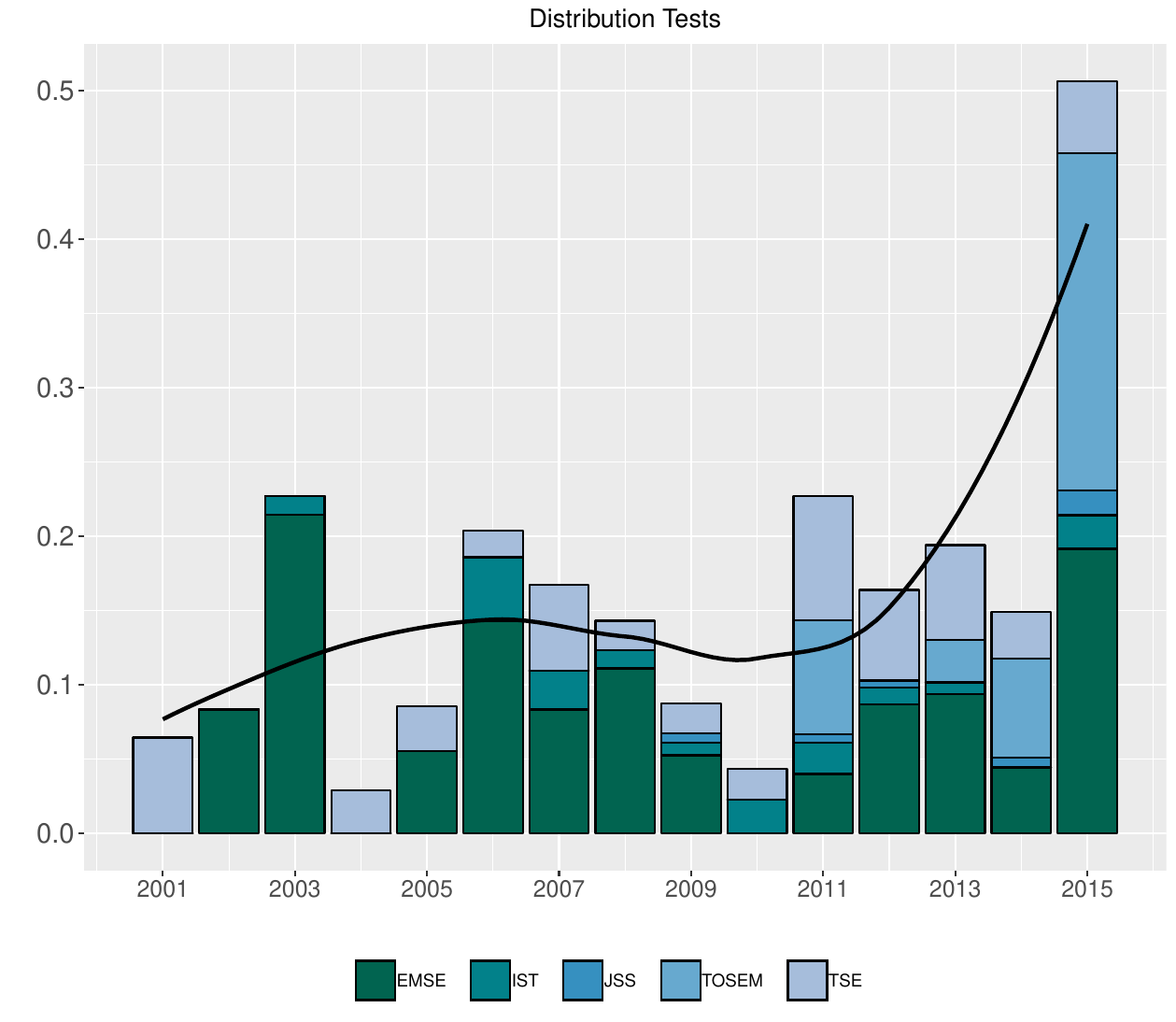}
    \caption{The use of distribution tests. One can see that after 2011 the amount of reported tests increases, even though they are still low (e.g., in 2015 all journals combined reached a score of 0.5, when the maximum in the scale is 5). }
    \label{fig:distTest}
\end{figure}

Normality tests include the Shapiro-Wilk and Kolmogorov-Smirnov tests; the latter is a more general technique that compares a sample with a reference probability distribution. A less analytic way of testing for normality is by plotting the normal quantiles against the sample quantiles (i.e., a normal probability plot) or by plotting the frequency against the sample data (i.e., frequency histograms) and, thus, visually assess the normality~\citep{ghasemi2012normality}. Figure~\ref{fig:distTest} provides a view of how distribution tests have been used over the years. Our extraction reveal that out of all matches for distribution tests (102) the most common tests are the Kolmogorov-Smirnov (45\slash 102) and Shapiro-Wilk (38\slash 102) tests.

\subsection{Parametric and nonparametric tests}\label{subsec:res:param_nonparam}
Parametric tests are based on assumptions regarding both the distribution of data and measurement scales used. On the other hand, nonparametric tests are popular because they make fewer assumptions than parametric tests, as known since the 1950s~\citep{anderson1961scales}. Surprisingly, studies in psychometrics have shown that parametric tests are surprisingly robust against lack of normality and equal variances (also called equinormality), with two important exceptions: one-tailed tests and tests with considerably different sample sizes for the different groups~\citep{boneau1960effects}.

\begin{figure}
    \centering
    \includegraphics[width=\textwidth]{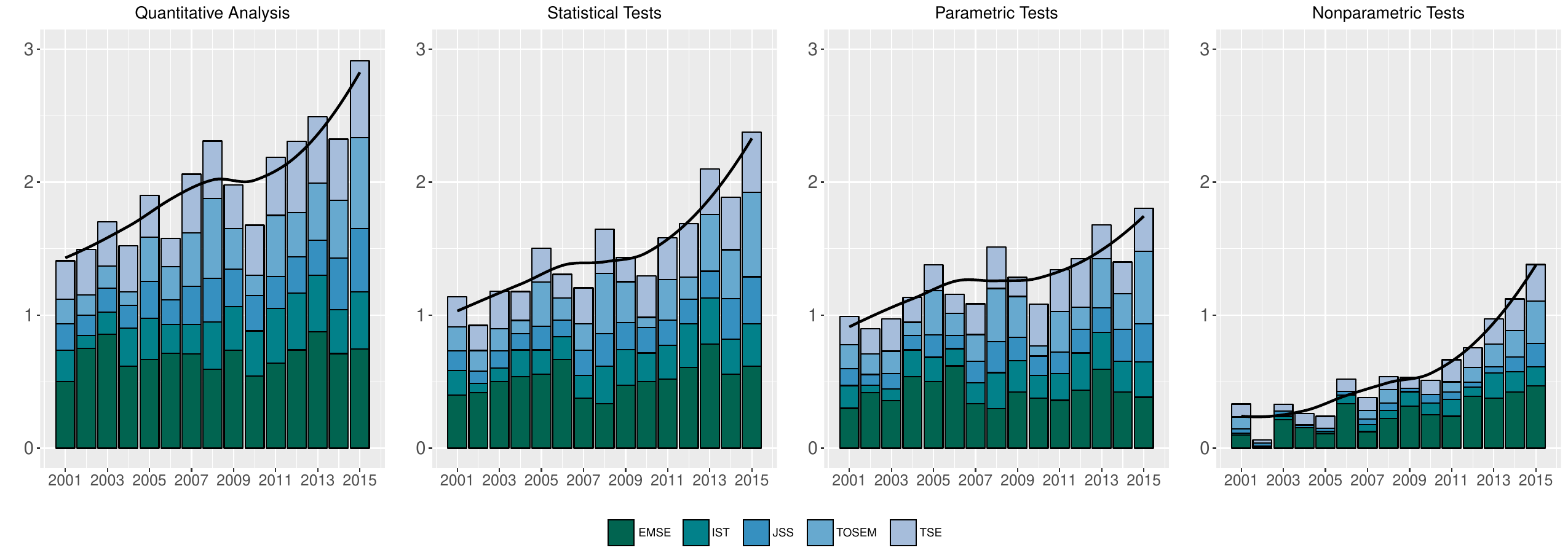}
    \caption{The $y$-axis in each chart is the normalization of ratings of the number of papers where we found positive evidence. Notice that the scale on the $y$-axis is still lower than the maximum ratio (5). The thick line is a local regression (loess) of the data.}
    \label{fig:combined1}
\end{figure}

In the case of software engineering,~\citet{ArcuriB2011hitchhiker} argue that data from more quantitatively focused studies (e.g., data that are not from human research subjects) provide data for which parametric tests are seldom applicable~\citep{ArcuriB2011hitchhiker}. Figure~\ref{fig:combined1} suggests a trend over the years of increased use of quantitative analysis, statistical tests, and parametric and nonparametric tests.

In particular, we see a strong increase in the usage of nonparametric statistics. Since~\citet{ArcuriB2011hitchhiker} published their paper on the usage of nonparametric statistics in 2011 a general increase of nonparametric statistics is visible. We found 504 papers using nonparametric tests, where the three most common nonparametric tests we found in ESE papers are Mann-Whitney $U$ test (202\slash 504) and its variations, the Kruskal-Wallis (83\slash 504) test, the $\chi^2$ (73\slash 504) test.

Nonetheless, parametric tests are still prevalent in our analysis being matched in 1171 papers. The three most used tests are: $t$-test (741\slash 1171), ANOVA (497\slash 1171) and $F$-test (136\slash 1171). Note that different tests (e.g., $t$-test and ANOVA) can be matched in the same paper when researchers have several dependent variables to analyse.


\subsection{Type I errors and how to avoid them}\label{subsec:res:TypeI}

\begin{figure}
    \centering
    \includegraphics[width=\textwidth]{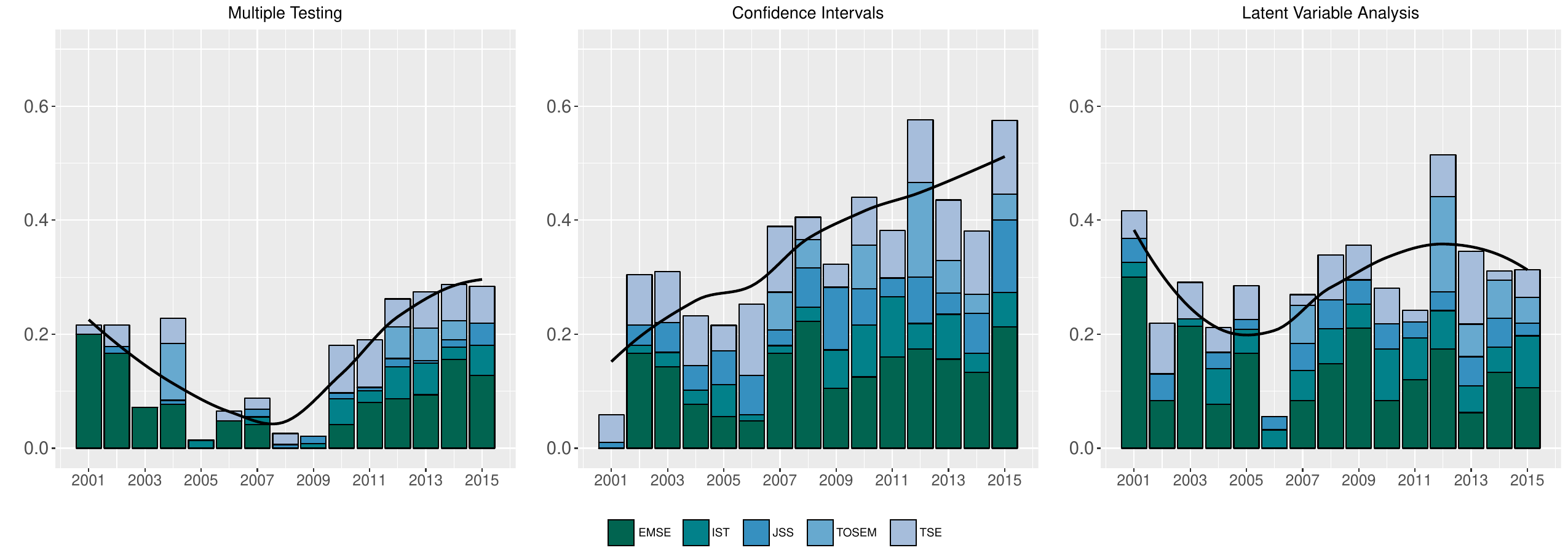}
    \caption{Usage of multiple test correction, confidence intervals and latent variable analysis in ESE over the years. The thick line is a local regression of the data. Note values on the $y$-axis range between zero and 1, so the trends shown are insightful but we show only a part of the scale (i.e., between 0 and 0.6).}
    \label{fig:combined2}
\end{figure}

A statistical test leads to a Type I error whenever the null hypothesis is rejected when it is true~\citep{pollard1987probability}. The probability of such an error is often denoted $\alpha$, and the probability of being correct is $1 - \alpha$. In turn, a Type II error is instead when we fail to reject the null hypothesis when the alternative is true, and the probability of such an error is often denoted $\beta$ (see Subsection~\ref{subsec:res:pwr}). 

There is a trade-off between the $\alpha$ and $\beta$ probabilities given a specific sample size. \citet{Dyba2006power} showed that the software engineering field tends to accept probability of error substantially higher than the standards of other sciences, which they suggested can be improved by more careful attention to the adequacy of sample sizes and research designs. Concerning $\alpha$, \textit{multiple testing} is a pitfall that underestimates this probability, which can be dealt with by adjusting $p$-values (e.g., by doing a Bonferroni correction which divides the obtained $p$-values by the number of statistical tests conducted)~\citep{benjamini1995controlling}. 

Our data reveals that only 125 papers are corrections for multiple testing, which is a very small number compared to the number of papers using parametric (1171) and nonparametric tests (504). Additionally, the papers tend to rely on a similar set of tests to analyse data (e.g., $t$-test and Mann-Whitney $U$). The most common techniques for controlling for multiple testing pitfall in empirical software engineering are Bonferroni (106\slash 125), Bonferroni-Holm (51\slash 125) and Tukey's range test (15\slash 125).

From an ESE researcher perspective, this raises two risks: many researchers are $i$) unaware of the multiple testing pitfall; or $ii$) neglect the existing negative results  (often called `the file drawer bias'). Ignoring to address both risks could mean that the published results are, in fact, Type I errors~\citep{Ioannidis05false,rosenthal1979file}. Nonetheless, it is reassuring to see the correction for multiple testing being used in ESE (Figure~\ref{fig:combined2} (Multiple testing)), even though the data indicates it could be used more often.\footnote{Note here that this is a preliminary assessment since $\alpha_K < 2/3$.} We further elaborate on this in Subsection~\ref{subsec:smm:stattests}.

\subsection{Confidence intervals}\label{subsec:res:CI}
Interval estimates are parameter estimates that include a sampling uncertainty. The most common one is the confidence interval (CI), which is an interval that contains the true parameter values in some known proportion of repeated samples, on average. The procedure computes two numbers that define an interval that contains the real parameter \textit{in the long run} with a certain probability. 

Figure~\ref{fig:combined2} (Confidence intervals) indicates that the use of CIs in our field is low (369 papers in total) but has a slight upward trend over the years. It is important to note that the implication ``if the probability that a random interval contains the true value is $x$\%, then the plausibility or probability that a particular observed interval contains the true value is $x$\%; or, alternatively, we can have $x$\% confidence that the observed interval contains the true value'' is unsound, as~\citet{MoreyHRLW2016CI} writes.

\subsection{Latent variable analysis}\label{subsec:res:lva}
Latent variable analysis (LVA) is a mathematical technique that assumes that there are hidden underlying variables that can not be directly observed. Such variables are, instead, inferred from other variables that can be observed (or directly measured). The main reasoning behind the mathematical model is that each observed variable expresses some combination of latent variables.
Preclsely, if there are $p$ observed variables $X_{1},X_{2},\ldots,X_{p}$ and $m$ latent variables $F_{1},F_{2},\ldots,F_{m}$, then $X_{j} = a_{j1}F_{1} + a_{j2}F_{2} +\cdots+ a_{jm}F_{m} + e_{j}$ where $j = 1,2,\ldots,p$, $a_{jk}$ is the `factor loading' of the $j$th variable on the $k$th variable, and $e_j$ is an error term. Each factor loading $a_{jk}s$ tells us how much  variable $F_k$ contributes to latent factor $X_j$. There are different estimation techniques for latent variable analysis; if the variance of the error terms is included, the resulting analysis is called a principle component analysis since we use all variance to find latent variables. If the error variance is excluded, it is often called a principle axis factor extraction~\citep{fabrigar}. 

In the context of building prediction models for software, factor analysis seems to be recommended~\citep{butte}. When it comes to human factors research in software engineering, the use of latent variable analysis seems a bit more scarce but recommended when, e.g., validating questionnaires~\citep{gren2016useful}. In Figure~\ref{fig:combined2} (Latent variable analysis) we see that the usage of latent variable analysis has been low (272 papers) in ESE over the years. The most common techniques are factor analysis (150\slash 272), principal component analysis (133\slash 272) and, far below, structural equation modeling (47\slash 272).

\subsection{Practical significance and effect sizes}\label{subsec:res:es}

\begin{figure}
    \centering
    \includegraphics[width=\textwidth]{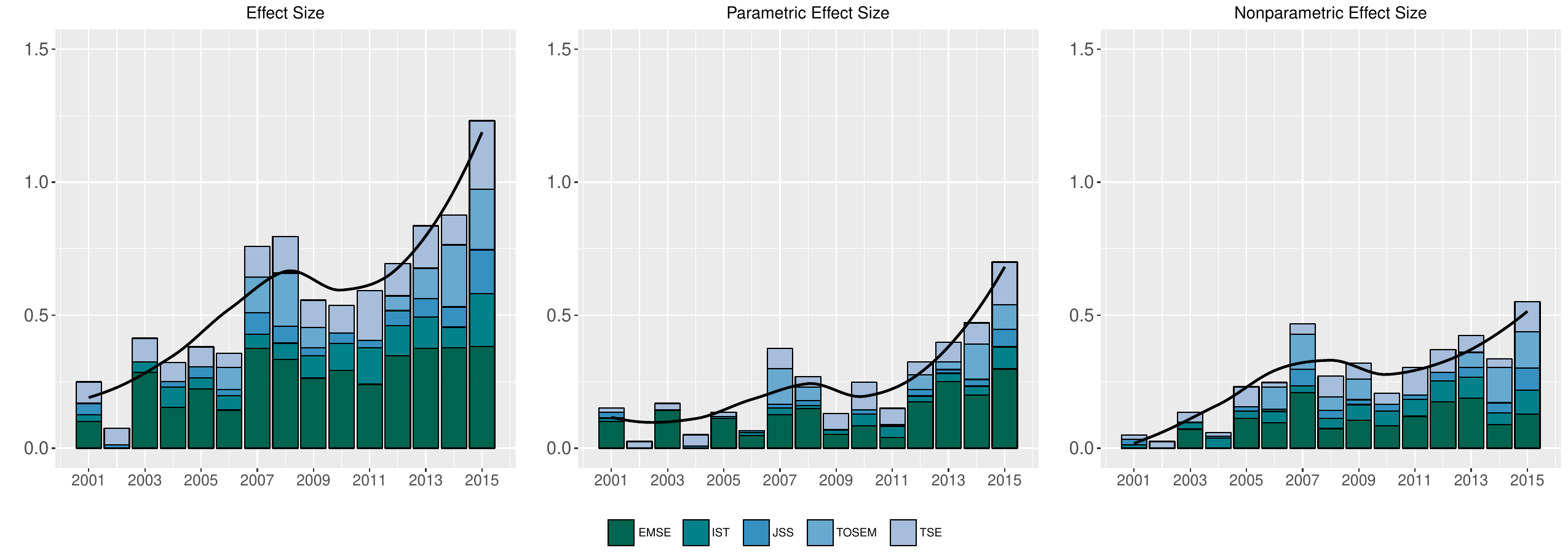}
    \caption{The left plot shows the usage of effect size statistics in general, while the other bar plots show the usage for parametric and nonparametric effect size statistics, respectively. The thick lines are local regressions.}
    \label{fig:combinedES}
\end{figure}

As an initial remedy to the strong limitations of null-hypothesis statistical testing, many fields have used effect sizes on significant results (see, e.g.,~\citep{becker2005potential, nakagawa2007effect}). There are numerous ways of calculating effect sizes depending on the research design. Two common families of effect sizes are standardized mean differences and correlation coefficients (based on variance explained, which can also be extended to multiple regression by the use of $R^2$)~\citep{cohen1992power}; knowing the magnitude of an effect, i.e., to what degree the null hypothesis is believed to be false, is an estimate needed in \emph{a priori} power analysis (see Subsection~\ref{subsec:res:pwr}).

There are also parametric and nonparametric, as well as standardized and non-standardized, effect sizes. The nonparametric ones do not assume any distribution, which makes them useful for non-normal data. Another family of effect sizes, called common language effect sizes, are effect sizes based on probabilities instead of estimates of size \emph{per se}. Such probabilities of effect magnitude have been advocated in software engineering lately, e.g., the Vargha-Delaney $\hat{A}_{12}$~\citep{ArcuriB2011hitchhiker}.

In a systematic review on effect sizes in software engineering from 2007,~\citet{Kampenes07es} showed that 29\% of the experiments reported effect sizes during 1993--2002 and the size of standardized effect sizes computed from the reviewed experiments were equal to observations in psychology studies and slightly larger than standard conventions in behavioral science.

Figure~\ref{fig:combinedES} illustrates how effect size statistics have been used over the years, in general, and for nonparametric\slash parametric, in particular. A total of 513 papers use effect sizes and both categories have had a positive trend. The most common effect size statistics we extracted are: Pearson's correlation (184\slash 513), Spearman's $\rho$ (104\slash 513) and Cliff's $d$ (63\slash 513). Several other effect size measures were identified, such as Odds Ratio, and the nonparametric $\hat{A}_{12}$. Note that, in some cases, extracting effect size information is can be difficult since researchers use a variety of methods to measure effect size. For instance, correlation coefficients (e.g., Pearson's correlation and $R^2$) have been used as effect size based on the variance of paired data~\cite{Maalej2013}. Researchers can also use coefficients of a multiple linear regression model to measures of effect size. However, such method is not standardized, hence being difficult to capture with our semi-automated extraction. Consequently, our analysis potentially skipped papers that discuss effect size.\footnote{Our tool detected 144 papers in total discussing any regression analysis, such that a subset of those papers are likely discussing effect size as well.} Since we already see a positive trend with the current set of papers, refining the tool to identify the false negatives would only increase the trends.

An implication of the positive trend are the effects on practical significance. Researchers often use effect size measures to discuss practical significance aimed at informing practitioners of the benefits of their proposed techniques. In other words, they aim to measure the magnitude of a statistically significant different in terms of its relevance in practice. However, we argue that discussing effects size is only a first step towards that goal. Most practitioners are not familiar with the nuances between different techniques used to convey this magnitude.

For instance, \citet{Tantithamthavorn2018icse-seip} report that using regression coefficients to interpret a model can be misleading and \citet{Furia2018bayesian} state that ``effect sizes can be surprisingly tricky to interpret correctly in terms of practically significant quantities''.\footnote{Authors suggest, in their context, using ANOVA Type-II/III test} Note that, reporting effect size is just one aspect of practical significance that must also consider a practitioner's perspective and their domain and, finally, be reported accordingly.

Practitioners are mostly concerned with the costs and challenges in adopting the techniques or the ROI that it provides, as opposed to generalization of results~\citep{Tantithamthavorn2018icse-seip}. If we want industry to introduce a certain technique\slash process\slash method, then providing an argument such as ``$X$ is significantly better than $Y$, with an effect size of 0.6'' is not so compelling compared to ``Training one staff member will require 2 hours; by using our algorithm  we will cut 6 hours of manual work for each case you handle.'' Consequently, the ESE community can benefit from more explicit discussions of practical significance (the latter example) further encouraging technology transfer of techniques.

\section{The conceptual model for statistical analysis workflow}\label{sec:ssm}
\begin{figure}
    \centering
    \includegraphics[width=0.8\textwidth]{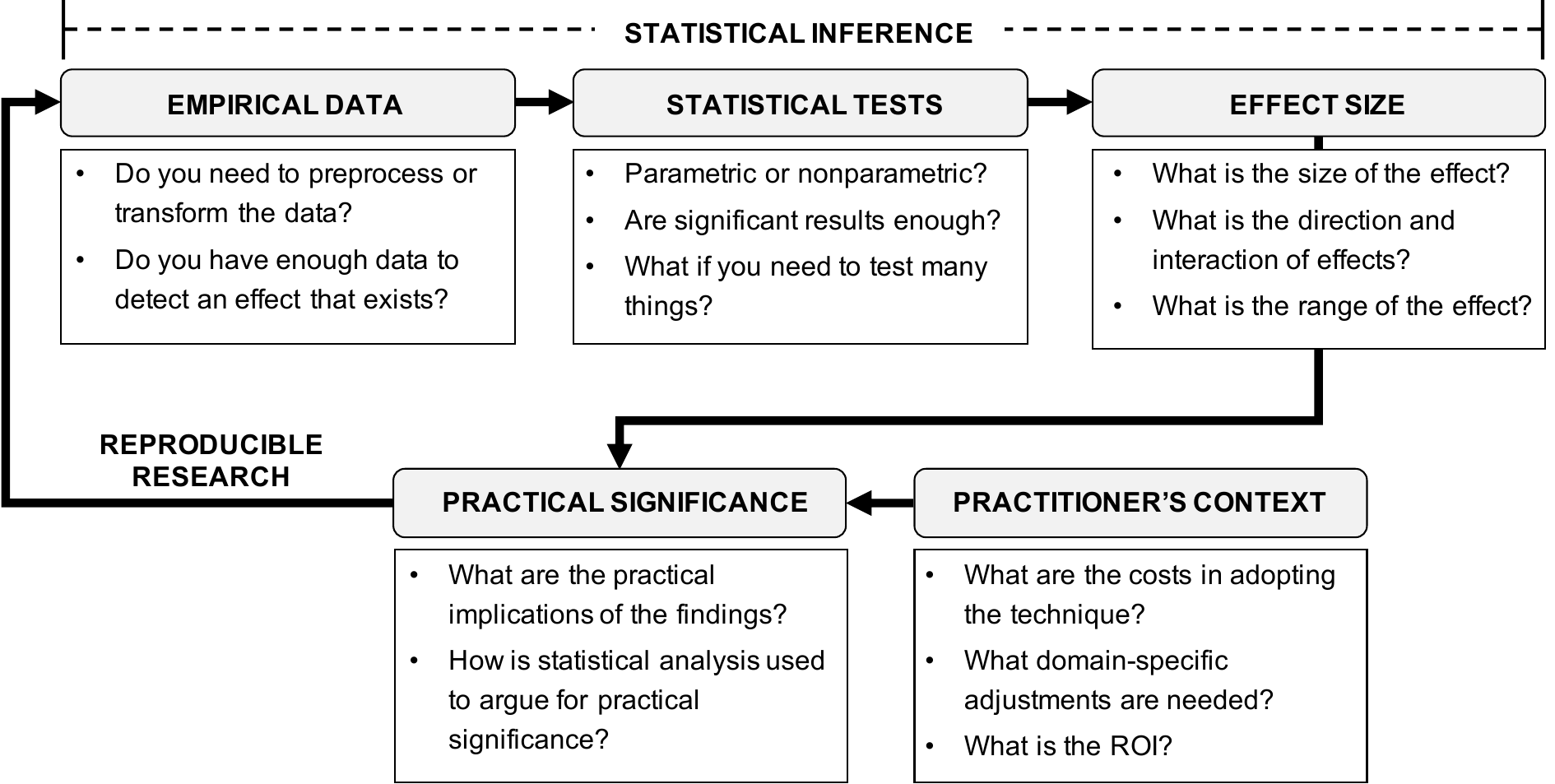}
    \caption{The statistical analysis workflow model for empirical software engineering research. Note that this does not show other, critical steps in the overall research process such as the choice of hypothesis or the design of the study itself and its data collection methods. Furthermore, the specific questions within each step will have to evolve as more advanced and mature statistical analysis techniques are added or starting to be used (e.g. Bayesian Data Analysis~\cite{Furia2018bayesian} etc).}
    \label{fig:smm}
\end{figure}

The results from the previous section provide us with input needed to analyze current statistical usage in ESE research and point to how we can strengthen the usage of robust statistical methods to better ground our discussions on practical significance. To this end, we present our conceptual model for statistical workflows in Figure~\ref{fig:smm}.

A conceptual model should guide a software engineering practitioner or researcher into thinking about the choices they make when conducting statistical analysis. Particularly, what type of questions should they answer and how the answers connect with each other. Ultimately, a full statistical argument usually requires to link the final analysis to the empirical data collected in the beginning.

Statistical analysis for an applied engineering science is a tool and not a goal in itself. Based on our findings on effect size and practical significance, researchers and practitioners are interested in knowing whether the data indicates a real effect, i.e., can repeatedly be found in related and similar situations, and that this preferably has practical consequences. Negative results are also of value since they help in deciding what might not be worth exploring further. But the value of negative results is in how it helps us re-design other studies and go back to find real and robust effects with practical significance. 

Each subsection below covers a particular aspect of our model in connection to the main practices and trends seen in the previous section. This section ends on a short note suggesting complementary statistical approaches that could be used to better support arguments of practical significance of research findings.

\subsection{Empirical data}\label{subsec:smm:data}
The key question when it comes to empirical data is \textit{if we have enough, so that we can detect a real effect if there is one}. Knowing the type of data is also important since it dictates which statistical analysis methods are appropriate and, thus, which assumptions may need to be fulfilled. For instance, our findings reveal that most papers do not employ methods to verify whether there is enough data to detect an effect (e.g., power analysis). We suggest the following questions that researchers should answer when starting analysis of collected data.

\paragraph{Do you need to preprocess or transform the data?} Note that statistical approaches can often require, or benefit from, transforming the data before analysis, e.g., through a log transformation (see~\citep{NeumannHP2015V-D} for a concrete example). Researchers should, nonetheless, explicitly report any transformations performed in the data and provide it in both the raw and processed forms. That allows others to verify the results or try alternative analysis methods. In particular, this encourages researches to criticise and improve of the preprocessing method itself~\citep{LarssonLT14}.

\paragraph{Do you have enough data to detect an effect that exists?} A robust statistical analysis should consider the sample size and do power analysis. As we discussed earlier,~\citet{Dyba2006power} have shown that the sample sizes we use in ESE are often not sufficient; we attribute it to two reasons: $i$) convenience sampling and, $ii$) lack of \emph{a priori} power analysis. Convenience sampling is often a consequence of limited access to companies or large data sets. Reproducible research practices, and repositories of dataset, such as PROMISE~\citep{PROMISE_2005}, can address this challenge on a shorter term. Several researchers emphasize the importance of enabling access to representative and relevant data~\cite{Tantithamthavorn2018icse-seip,GomesTM15,GonzalesB12,Gomez2010} and explicitly fostering power analysis can leverage this relevance.

In addition to catering for sample size, researchers should keep measurement errors as low as possible. \citet{gelman18failure} reports that we should focus on reducing measurement errors and move from between-person to a within-person design when possible. For instance, reducing the measurement error by a factor of two is comparable to multiplying the sample size by a factor of four~\citep{gelman18failure}. Regarding \emph{post hoc} power analysis we note that it has been regarded as controversial~\citep{Thomas97pwr} and many statisticians advise against its use. 

In short, power analysis contributes to the robustness of the analysis at the level of the empirical data. In order to foster such robustness, we summarize the following guidelines:

\begin{mdframed}[style=style1]
\begin{itemize}
    \item[G0.1:] A robust statistical analysis should consider \emph{a priori} power analysis, which is widely supported by modern statistical tools.\footnote{For instance, in \textsf{R} the \textsf{pwr} package~\citep{pwr} provides power analysis according to Cohen's proposals~\citep{cohen1988spa}.}
    \item[G0.2:] Researchers should consider not only related work but \textit{explicitly reasoning} about the strength of previous research (e.g., existing datasets).
\end{itemize}
\end{mdframed}

\subsection{Descriptive statistics}\label{subsec:smm:descr}
Before selecting and applying specific statistical analysis, researchers should understand the data and what kind of variation it exhibits. Even though descriptive statistics alone do not entail statistical robustness, they summarize relevant information about the investigated phenomena. 

Our results indicate that most empirical papers report descriptive statistics. Discussion in terms of means, medians, standard deviations, or visualizations and graphs (e.g., box plots), can trigger insights and act as a first checkpoint before moving towards more complex statistical methods. Moreover, discrepancies at this checkpoint indicate that researchers should revisit the previous stage and check their empirical data. To this end, we suggest the following question to guide researchers in reporting their descriptive statistics:

\paragraph{What are the key properties of the data?} Researchers should then be mindful to report at least: one central tendency measure (mean, median or mode), one dispersion (standard-deviation, variance or percentiles) and plot the shape of the data (skewness). Such key numerical summary statistics can reveal problems or pitfalls with the data.

\paragraph{What does the data look like?} The most commonly used graphs are box plots, but kernel density plots are a better choice since they are not as as sensitive to bin size and provide more detailed information such as skewness of the data~\citep{Kitchenham2017robust}. Other examples of recommended plots are scatterplots and beehive plots since they show individual data points, hence increase understandability.

\paragraph{Which assumptions are supported by the data?} Even though plots can be used to judge the distribution of data (e.g., normal probability plots), researchers should use statistical tests such as goodness-of-fit and distribution tests (e.g., tests of normality) given that these methods are less subjective and error-prone~\citep{ArcuriB2011hitchhiker}. In fact, our results indicate a positive trend in using distribution tests to support assumption of the data. Consequently, researchers will have more reliable evidence to decide whether parametric or nonparametric statistical analysis is called for.

Our analysis reveals that the way researchers often use Kolmogorov-Smirnov indicate that Lilliefors correction~\citep{Lilliefors67} is not widely used (10 out of 102 papers). Lilliefors correction does not specify the expected value and variance of the distribution, but the rule is still out if this is a good thing~\citep{YapS11k-s}. Additionally, several studies have shown~\citep{Razali11a-d, YapS11k-s} that the Shapiro-Wilk or Anderson-Darling~\citep{AndersonD1952a-d} tests are preferable to Kolmogorov-Smirnov, due to having greater statistical power.
 
Given our discussion above, we summarize the following guidelines for using descriptive statistics and investigating assumptions about the data:

\begin{mdframed}[style=style1]
\begin{itemize}
    \item[G1.1:] Researchers should report on different types of descriptive statistics in their data, including: central tendecy, dispersion and shape. Such information can act as a first checkpoint of the statistical analysis.
    \item[G1.2:] Plots and visualization of data should enhance and complement understanding of the descriptive statistics. Examples of preferred plots are kernel density plots (emphasizing dispersion in terms of shape of the data), scatterplots and beehive plots (emphasizing dispersion in terms of individual datapoints).
    \item[G1.3:] Researchers should use distribution tests to support assumptions about the data (e.g., normality), instead of simple visual analysis of density plots and histograms. We recommend Shapiro-Wilk or Anderson-Darling instead of the widely used Kolmogorov-Smirnov.
\end{itemize}
\end{mdframed}
 
\subsection{Statistical testing}\label{subsec:smm:stattests}
Statistical testing is a decisive tool to determine whether the data collected point to a difference between different levels of a factor, or even between factors themselves. Our findings reveal that ESE relies primarily in frequentist approaches to determine the significance of this difference. Literature offers a variety of statistical tests, employed throughout many ESE papers. 

For instance, we identified, between 2001--2015, a total of 1,380 papers using any of more than 10 statistical tests analysed. However, all tests do not have the same applicability. Some tests are more permissive to different types of data (e.g., dispersion, shape, scales) than others. Choosing the wrong test to measure whether there is a significant difference between treatments can lead to severe conclusion validity threats, implicating, even, invalid or wrong conclusions~\citep{WohlinRHOR2012exp}. Below, we create a list of questions authors can use when choosing statistical tests.

\paragraph{Parametric or nonparametric models?} \citet{ArcuriB2011hitchhiker} state, and our data confirms (Figure~\ref{fig:combined2} in Sectoin~\ref{sec:results}), that traditional parametric statistical tests are the most popular choice in ESE. However, researchers often neglect to check relevant assumptions that can affect the conclusions of a parametric tests, such as having data following a normal distribution. For instance, even though our data reveals that 1,380 papers use any type of statistical tests, our tool only detected 102 tests (less than 10\%) that report on a statistical test. Particularly, 72 papers using any parametric test report on at least one distribution test for normality, as opposed to 63 papers reporting on nonparametric tests. 

Nonetheless, the remaining papers 1,278 papers using statistical tests, but neglecting to report distribution tests should be, if possible, re-analysed in future work in order to identify possible conclusion validity threats not raised by corresponding authors. The choice of tests is up to the research and the constraints from the data, but we agree with~\citet{ArcuriB2011hitchhiker} that nonparametric tests are preferable, given their fewer assumptions.

\paragraph{Are significant results enough?} The main tool used to convey a statistically significant difference in a statistical test is the comparison between a $p$-value returned from the test and a threshold value denoted significance level (or $\alpha$). The choice of values for $\alpha$ and misuse of $p$-values have raised a lot of discussions in statistics and is currently under scrutiny~\cite{wasserstein2016asa,briggs2017substitute}. Moreover, this discussion is moving now to ESE~\citep{Krishna2018smells_analytics,Furia2018bayesian}, as many researchers do not motivate the choice of significant value when reporting statistically significant difference. Instead, authors choose arbitrary values (e.g., $\alpha = 0.05$ or $\alpha = 0.01$) without providing any reasoning about this choice~\citep{Krishna2018smells_analytics}. Researchers can motivate the choice of $\alpha$ in terms of the probability of a Type-I error. Note that, just because a test shows statistically significant results that does not necessarily imply evidence of a real effect~\citep{Thompson02NHST}.

For instance, by using $\alpha = 0.05$ as a threshold, researchers accept up to a 5\% probability of rejecting a null hypothesis, when the hypothesis is true (i.e., they should not have rejected it in the first place). Null hypothesis are designed so that the investigated ojbects of study are not different from each other. In other words, for a software project that means taking the chance of adopting a new\slash different technique when, in fact, the new technique would not provide any significant benefit. In practice, practitioners could, perhaps, be willing to take higher (or smaller) chances in adopting such new techniques, if the probability that there is no difference between techniques do not result in a severe loss. 

In summary, the choice of $\alpha$ to convey statistical significance should be motivated and, ideally, one should consider a practitioner's opinion on the acceptable thresholds for such probabilities values. Over time we also think that alternatives to the use of traditional hypothesis testing and the use of $p$-values will be preferable. However, there is not yet a consensus, even among statisticians, on which alternatives to use so, in the meantime, ESE researchers should improve the way they use hypothesis testing if they decide to use it.

\paragraph{What if you need to test many things?} Our analysis reveals that many papers do not correct for multiple tests at all (only a total of 125 papers report usage of multiple test corrections). In fact, \citet{Kitchenham2019practices} report similar risks when performing pre-testing (e.g., normality tests) that may also be affected by multiple tests, hence leading to uncontrolled Type 1 and Type 2 error rates.

Many modern statistical software provide the option to including such corrections when performing pairwise statistical testing of different levels within a factor.\footnote{The function \texttt{p.adjust()} in \textsf{R} supports eight different methods for corrections: \url{https://bit.ly/2XgImdn}} Given the amount of different tests available in literature, we recommend that Bonferroni-Holm should be used over Bonferroni since it offers greater $\beta$ (i.e., Type II errors). Alternatively, researchers can use Benjamini-Hochberg since there is evidence that it might be optimal despite having additional assumptions on the data. Our results on multiple testing extraction (Section~\ref{subsec:res:TypeI}) show however, that more papers report on Bonferroni, as opposed to Bonferroni-Holm.

In summary, we propose the following guidelines for researchers aiming to use statistical tests:

\begin{mdframed}[style=style1]
\begin{itemize}
    \item[G2.1:] The choice of statistical tests have severe impact on conclusion validity~\cite{WohlinRHOR2012exp}, hence researchers should understand the assumptions and constrains inherent to the different statistical tests in their toolkit.
    \item[G2.2:] Researchers should perform distribution tests (e.g., normality tests) on their data before using parametric tests~\citep{ArcuriB2011hitchhiker}. Neglecting to check the assumptions related with the test can lead to wrong results, hence nonparametric tests are preferred since they have less assumptions and less constraints concerning the data.
    \item[G2.3:] Avoid using arbitrary values for $\alpha$. Considering that $\alpha$ acts as a threshold for $p$-values, researchers should motivate the choice of such a threshold in terms of the consequences behind a Type I error in their specific context (i.e., a practitioner's perspective).
    \item[G2.4:] Researchers should use and report multiple-testing when using frequentist approaches in order to correct for the Type I and Type II error rates.
    \item[G2.5:] Researchers should inform themselves about alternatives to $p$-values and hypothesis testing (as currently being discussed and questioned within the area of statistics itself) and evolve their methods as a wider consensus is reached.
\end{itemize}
\end{mdframed}




\subsection{Effect sizes}
After detecting a statistically significant effect in a frequentist approach, researchers use effect size measures to determine how large such an effect is. Our findings reveal an increased usage of effect size statistics (Subsection~\ref{subsec:res:es}) but no specific trend to the specific types of measures (i.e., parametric and nonparametric effect sizes), even tough there is higher increase in the usage of nonparametric statistical tests instead.

Note that an increased usage of nonparametric testing also includes other types of effect size measurement, i.e., $\hat{A}_{12}$. However, it can be confusing to call $\hat{A}_{12}$ an ``effect size'' measure, since it is essentially a probability estimate (how likely is it that an investigated technique $X$ is better than technique $Y$). Such confusion can lead to misunderstanding in interpreting the results of $\hat{A}_{12}$ hence introducing severe conclusion validity threats.

For instance, interpreting that an $\hat{A}_{12} = 0.8$ indicates that ``technique X is 80\% better than another'' is wrong. Instead, $\hat{A}_{12} = 0.8$ simply conveys how often one is better if we repeatedly compare them (i.e., that 80\% of the time, technique $X$ is better than technique $Y$). Despite this possibility for misinterpretation, the $\hat{A}_{12}$ statistics is still easier to understand than parametric effect sizes such as Cohen's $d$ with its ranges (e.g., small, medium or large effect) that are subjective. Nonetheless, the choice between parametric and nonparametric test on the previous stage of our model should also guide the selection the type of effect size measure used.

\paragraph{What is the direction and interaction of effects?} While the traditional statistical tests can tell if an effect is likely, they are not enough for understanding the strength of interaction of multiple factors (i.e., confounding variables). In other words, researchers should be mindful that the effect of one independent variable on an outcome may be dependent to the state of another independent variable.

Regression analysis can help build more complex models, and its extension to generalized linear models makes regression applicable regardless of the underlying data generation process~\citep{faraway2006elm}. Moreover, analysing the coefficients of these variables have also been a practice to evaluate effect size. Even when we build a regression model, we cannot conclude that the predictor variables necessarily \textit{cause} what comes out of the model. In recent years there has been a lot of progress on extending statistics to the analysis of causality, i.e., not only that two variables are correlated but which one is causing changes in the other. For instance, researchers can use casual analysis to identify the direction of the causality~\citep{schoelkopf17ci,Greenland1999dag_pearl}. 

\paragraph{What is the range of the effect?} Our data shows that the usage of confidence intervals (CIs), to show the likely range of the size of a variable or effect, has remained steady in ESE over the years. About $\sim$20\% of the manually reviewed 2015 studies use CIs in some form. From a frequentist perspective, the reporting of CIs is still considered a good practice, however there is a common pitfall when interpreting confidence intervals. The confidence coefficient of the interval (e.g., 95 \%) is thought to index the plausibility that the true parameter is included in the interval, which it is not. Moreover, the width of confidence intervals is thought to index the precision of an estimate, which it is not~\citep{MoreyHRLW2016CI}. As an alternative, researchers can use Bayesian credible intervals that shows the plausibility that the true parameter is included in the interval, hence being more informative than confidence intervals.

In short, we suggest the following guidelines when analysing effect sizes:

\begin{mdframed}[style=style1]
\begin{itemize}
    \item[G3.1:] Using parametric or nonparametric statistical tests should guide researchers to parametric or nonparametric effect sizes, respectively.
    \item[G3.2:] When using common language effect size statistics (e.g., $\hat{A}_{12}$), the interpretations should be carefully reported from a \emph{probabilistic} point of view.
    \item[G3.3:] Report on the direction of your causality. Given their simplicity, DAGs help other researchers understand better the dependencies between variables in an experiment~\cite{Greenland1999dag_pearl}. Additionally, researchers can use, for instance, additive noise models to identify the causality direction of their variables~\citep{Peters2014causal}.
    \item[G3.4:] When reporting confidence intervals, discuss the pitfalls about the confidence coefficient and its interpretation. Alternatively, researchers can use Bayesian credible intervals. 
\end{itemize}
\end{mdframed}

\subsection{Practical significance}
The workflow used in the statistical methods discussed in the previous steps of our model should lead to an analysis on practical impact of the findings. We argue that there are two orthogonal aspects when analysing and reporting practical significance: $i$) proper choice and usage of statistical methods (discussed in the previous subsections), and $ii$) the context provided by a practitioner (or other stakeholder) that intends to apply the proposed method or technique.

Proper choice and usage of the statistical analysis workflow must be at the basis of such a discussion as a measure for the validity and reliability of the evidence collected. Consequently, if we are not likely to see (i.e., statistically significant) a large enough effect (effect size of enough magnitude), then the evidence is not sufficient to support a claim that results have a practical significance. Nonetheless, results may still be relevant to practitioners in their specific context. In fact, \citet{Tantithamthavorn2018icse-seip} argue that enabling practitioners to interpret practical significance in terms of their own context is more relevant to the practical impact of a study rather than aiming for generalization alone.

Researchers should avoid implicit and vague arguments for practical significance based solely on the nature of the problem\slash context. (e.g., ``Technique $A$ helps reducing testing costs in industry''). Instead, researchers should be able to report the achieved statistical conclusions in terms of one or more contexts where the investigated technique should be applied. This allows both researchers and practitioners to identify which variables would be affected by the effect in the affected contexts. Variables are typically high-level observable constructs like cost, effort\slash time, or quality, but can also be more detailed metrics related to the top-level ones. 

During our manual reviews, we identified three ways of generally discussing practical significance: explicitly, implicitly or not at all. When discussing explicitly, authors should incorporate the evidence gathered during the statistical analysis. The following examples found during extraction illustrate these scenarios: 

\paragraph{Explicit (with evidence)} ``Each finding is summarized with one sentence, followed by a summary of the piece of evidence that supports the finding. We discuss how we interpreted the piece of evidence and present a list of the practical implications generated by the finding''~\citep{CeccatoMMNT15quote}. Practical significance is discussed explicitly and in direct connection to the results of the statistical analysis.

\paragraph{Explicit (without evidence)} ``The aforementioned results are valuable to practitioners, because they provide indications for testing and refactoring prioritization''~\citep{AmpatzogluCCA15quote} or ``The proposed approach can be utilized and applied to other software development companies''~\citep{garousi2015usage}. Even though the first statement might seem a bit more explicit compared to the second statement, neither has a clear connection to the output from statistical analysis.

\paragraph{Implicit} ``Price is the most important influence in app choice [\ldots], and users from some countries [are] more likely than others to be influenced by price when choosing apps (e.g., the United Kingdom three times more likely and Canada two times more likely)''~\citep{LimBKIH15quote}. The practical significance of the finding, even though it is not discussed, might lead to a company choosing a more aggressive differentiation of prices depending on the country.

In turn, in order to include the practitioner's context into the practical significance discussion, \citet{Torkar2018utility_model} suggest the use of cumulative prospect theory (CPT)~\citep{Kahneman1979cpt} in combination with the observable constructs of the experiment. For instance, cumulative prospect theory (CPT) represents how people judge risks and outcomes and handle risks of decision-making in a non-linear fashion. The risk values can be connected to the outcome of the statistical analysis, to support practitioners in deciding whether the investigated technique should be adopted~\citep{Torkar2018utility_model}.

In short, we suggest the following guidelines for analysing practical significance:

\begin{mdframed}[style=style1]
\begin{itemize}
    \item[G4.1:] Explicitly report the practical significance both to your statistical analysis and a practitioner's context. The results on effect size should be complemented\slash contrasted with contextual information such as costs for adopting a technique, complexity to understand or operate the techniques, technical debt, etc.
    \item[G4.2:] Connect the statistical analysis to a utility function that allows practitioners to interpret statistical results (e.g., $p$-values, effect sizes) in terms of risks of adopting the technique (e.g., using CPT)~\citep{Torkar2018utility_model}.
\end{itemize}
\end{mdframed}

\subsection{Levels of reproducibility}
In our conceptual model (Figure~\ref{fig:smm}), reproducible research allows the outcome of an empirical study to become empirical data for future empirical studies. Therefore, reproducibility practices must be addressed throughout all different steps of an experiment considering that each yields artifacts (e.g., processed data, charts, plots, discussions) that must be reported to the community. Current empirical guidelines emphasize the importance of packaging the research~\citep{Tantithamthavorn2018icse-seip,Krishna2018smells_analytics,Carver2014,Gomez2014,GonzalesB12,WohlinRHOR2012exp}, where all documentation, data, and apparatus, related to the study, must be made available so that other researchers can understand and reuse them. 

When designing a study, researchers need to be aware of existing data regarding the investigated causation, and how that data affects their investigated sample concerning a population. Therefore, searching and uploading data to open repositories such as PROMISE\footnote{\url{http://promise.site.uottawa.ca/SERepository/}} or Zenodo\footnote{\url{https://zenodo.org/}} should be a requirement for empirical studies. Otherwise, we risk to accumulate overlapping, yet unrelated, conclusions that could, instead, be aggregated towards generalization or insightful for practical significance.

Unfortunately, researchers are rarely provided with reproducible packages containing reusable information from existing studies in literature~\cite{Krishna2018smells_analytics,GomesTM15}. Availability, accessibility, and persistence of empirical data are important properties to assess reproducibility~\citep{GonzalesB12}. Current initiatives, such as the ACM badges for artefact reviews,~\footnote{\url{https://www.acm.org/publications/policies/artifact-review-badging}} enable researchers to systematically assess the level of data availability, hence reproducibility of studies.

In parallel to providing reproducible packages, researchers should also improve reporting of their results. Even though there are many trade-offs in using standardized paper structures,~\citet{Carver2014} suggest that reproduced studies include information on, at least, $i$) the original study, $ii$) the reproduced study, $iii$) comparison between the studies (e.g., what was consistent\slash inconsistent throughout both studies) and $iv$) conclusions across both studies. The replication done by \citet{Furia2018bayesian} is an example that follows this guideline. Moreover, \citet{Krishna2018smells_analytics} provide a list of seven ``smells'' that can be found in reported studies, such as including incomplete or too high-level description of constructs, and arbitrary experiment parameters devoid of justification.

In short, previously published ESE literature, by and large, recommends the guidelines below in order to address current limitations in empirical studies (sample size, hindered generalization, etc.) and help researchers to gauge a foundation to continuously evaluate empirical data.

\begin{mdframed}[style=style1]
\begin{itemize}
    \item[G5.1:] Report results consistently. \citet{Krishna2018smells_analytics} present a list of ``smells'' that can be identified when reporting empirical studies.
    \item[G5.2:] Use repositories of empirical data, since empirical data is the baseline for any empirical study and both using and contributing to such repositories strengthens the community in overcoming challenges with sample sizes and representative data.
    \item[G5.3:] Use systematic reviewing protocols, such as the ACM artefact badge, to evaluate reproducible research. Distinguish between the quality of the documented artefact, its availability, and the difficulties in achieving the same results (e.g., technical problems with scripts or tools).
\end{itemize}
\end{mdframed}


\subsection{Complementary statistical methods for ESE}\label{subsec:bayes}


Our data shows that ESE relies on different practices comprising a variety of statistical methods and tools to support researchers in validating their findings. On the other side, a subset of methods is predominant in most studies across the years (e.g., parametric tests, specific normality tests). There is a prevalence of frequentist approaches and, particularly, the reliance on hull hypothesis significance testing and a predominant subset of methods across the years (e.g., specific parametric tests and normality tests).

Several research fields have recently come to realize that the focus on $p$-values is not beneficial~\citep{Nuzzo14errors,TrafimowM15editorial,Woolston15pvalues,wasserstein2016asa}. One answer to this has been the introduction of effect size statistics~\citep{ArcuriB2011hitchhiker,becker2005potential,nakagawa2007effect}, but unfortunately, this does not fully address the issues with $p$-values (e.g., risks with Type I and II errors, choosing wrong tests, neglecting to check assumptions related to specific statistical tests). In this subsection we list several approaches that can be used to diversify and complement the statistical toolkits of ESE researchers. The goal is to inform researchers of strategies that addresses several of those pitfalls that are inherent to frequentist approaches. While our guidelines above can improve on current practices the approaches below have the potential to take statistical analysis in ESE even further. Thus, over time we argue both that use of these methods in ESE should and are likely to increase.

\noindent \textbf{Bayesian Data Ananlysis (BDA):}
As several authors have argued~\citep{GelmanCSDVR14BDA,McElreath15stat,robert2007bayesian}, Bayesian data analysis provides different advantages when conducting statistical analysis~\citep{Furia2018bayesian}. In general, the Bayesian data analysis workflow is more informative about the data and the outcome of the analysis when compared to frequentist approaches~\cite{GelmanCSDVR14BDA}. 

For instance, BDA requires the researcher to design a statistical model describing a phenomenon and, in the end, know the \textbf{uncertainty} of the parameters (e.g., factors) in that model, allowing researchers to interpret both whether there is an effect (statistical significance), but, more importantly, the size of this effect~\cite{GelmanCSDVR14BDA}. In contrast, most frequentist approaches provide a point estimate (i.e., a single $p$-value) that is compared to a threshold ($\alpha$) to obtain statistical significance and only then use a different measure to obtain the effect size.

Another advantage of BDA to ESE is the usage of \textit{priors} (i.e., existing data about a certain phenomena). Systematically reasoning about existing data is crucial for solving the replication challenges we have in software engineering research~\citep{Furia2018bayesian,Gomez2014,GonzalesB12,GomesTM15}; before these challenges get worse as in other disciplines~\citep{benjamin18crisis}. Priors can also allow the more systematic buildup of chains of supporting evidence and studies in SE and avoid that studies stand alone and has to be only qualitatively interpreted as a whole.

Advances in computational statistics have enabled researchers to use advanced BDA through platforms such as Stan.\footnote{\url{https://mc-stan.org/}} Consequently, BDA entails more involved computations and analysis when compared to a frequentist analysis. Nonetheless, BDA allows researchers to overcome numerous analysis pitfalls and obtain a more detailed analysis of the phenomena, a relevant introduction for ESE can be found in ~\citep{Furia2018bayesian}.

\noindent \textbf{Missing data:}
Missing data can be handled in two main ways. Either we delete data using one of three main approaches (listwise or pairwise deletion, and column deletion) or we impute new data. 

Concerning missing data, we conclude the matter is not new to the ESE community. \citet{liebchenS08dataquality} has pointed out that the community needs more research into ways of identifying and repairing noisy cases, and \citet{mockus2008mi} claims that the standard approach to handle missing data, i.e., remove cases of missing values (e.g., listwise deletion), is prevalent in ESE.

We argue that a more careful approach would be wise and avoid throwing data away. Instead, researchers should use modern imputation techniques to a larger extent. For instance, sequential regression multiple imputation has emerged as a principled method of dealing with missing data during the last decade~\cite{vanBuuren07mice}. Imputation can be used in combination with both frequentist and Bayesian analysis.\footnote{The R package supports imputation of missing data: \url{https://cran.r-project.org/web/packages/mice}} However, by using multiple imputation in a Bayesian framework, researchers obtain the uncertainty measures for each imputation. On the other hand, the computational effort increases and the analysis becomes more involved.

\noindent \textbf{Causality analysis:} Researchers should explicitly report on their scientific model involving the dependencies between variables and directions of causality. One can use directed acyclic graphs (DAGs) to this end, a concept refined by Pearl and others~\citep{pearl09causality}. Including such representations allows other researchers to criticise and reuse a study's scientific model, particularly when aiming for reproduction.

Instead of using the graphical approach of DAGs, researchers could use $do$-calculus to determine $d$-separation and numerical approaches to let the data build a DAG~\cite{schoelkopf17ci}. Nonetheless, causality analysis should become more common in many disciplines in the future, regardless of what statistical approach one takes, since they enable fundamental understanding about the phenomena and the empirical study itself.

\section{Discussion summary and threats to validity}\label{sec:threats}
Based on both manual and semi-automated review of a large set of empirical research papers in software engineering (SE) we have uncovered trends in their use of statistical analysis. By comparing this with existing guidelines on statistical analysis we presented a conceptual model as well as guidelines for the use of statistical analysis in empirical SE (ESE). This can help both researchers better analyse and report on their ESE findings as well as practitioners in judging such results. Below we discuss the relationships between our findings and how they relate to our research questions.\\

\noindent \textbf{RQ1: What are the main statistical methods used in ESE research, as indicated by the studies that are published in the main journals?}

Both of our review processes revealed a similar proportion of papers reporting statistical methods when considering the total pool of papers published in the corresponding sample within all selected journals. Specifically, our manual review revealed that 33.5\% of 2015 papers (161\slash 480) included a statistical method, whereas our extraction detected 36.1\% of papers (1,879\slash 5,196). Consequently, we cannot claim that the majority of papers report statistics, however, there is a positive trend when it comes to using different statistical methods. Particularly, our extraction indicates a focus on parametric tests (specifically $t$-test and ANOVA).

On the other hand, the number of papers that use distribution test (102) account for only 7\% of the number of papers using either parametric or non-parametric tests (1,380 papers). This is a strong indication that researchers neglect to verify underlying assumptions regarding data distribution required to properly use parametric tests. 

One can argue that the missing number of papers is attributed to false negatives matches from our tool, particularly if researchers use plots (e.g., histograms) to justify normality. We dispute that argument based on two premises: $i$) distribution tests are simpler to match, since they are keyword based (unlike, for instance, detecting effect size analysis with regression coefficients);\footnote{In addition to the tests names, the analyzer used in \texttt{sept} tries to match the use of expressions like ``normality test'' or ``check for normality'' during the extraction.} and $ii$) the gap is still too big, such that 93\% of the papers reporting statistical tests do not include checks for normality.\\

\noindent \textbf{RQ2: To what extent can we automatically extract usage of statistical methods from ESE literature?}

In order to enable analysis of a large pool of papers spanning through several years, we designed \texttt{sept} to match regions of the text of the papers with specific keywords and examples extracted during the first round of manual reviews. The design allows researchers to include more examples in each extractor file, or create their own using our proposed DSL.

Automated extraction is feasible and, our validation, shows that it is reasonably reliable. Note that extracting specific keywords, such as tests with specific names is easier but also tricky, since the matching can capture several false positives (as we illustrated in our validation when matching ``$t$ test'' with the term uni\textit{t test}). Researchers using \texttt{sept} can specify skip and negative matchers in each analyzer to avoid such cases.

On the other hand, extracting description of statistical analysis that rely on complex reasoning that spans across different parts of the paper requires a more robust algorithm, hence yielding false negatives. Nonetheless, we state that the purpose of \texttt{sept} is not to fully automate analysis of primary studies, rather our goal is to provide semi-automated tool support for checking properties of papers. An example of extension to \textit{sept} is to automatically detect violations of double-blind review processes in conferences, where names, emails or institutions of authors can be detected using analyzers.

Even with the current limitations, we were able to identify insightful trends and extract meaningful information about different statistical methods used in ESE across 15 years. A version of \texttt{sept} is available online\footnote{\url{https://hub.docker.com/r/robertfeldt/sept}}, but the sample of papers used cannot be made available given access constraints from publishers. However, all analyzers written for each extractor is available and detailed in the tool's documentation.\\

\noindent \textbf{RQ3: Are there any trends in the usage of statistical techniques in ESE?}

Based on our semi-automated extraction, there is an upward trend for using more statistical methods in general. Particularly, we identified positive trends for using nonparametric tests, distribution tests and effect sizes. On the other hand, the proportion of papers using such techniques in comparison to the pool of papers is still small, particularly when considering effect size, multiple testing, or distribution tests.\\

\noindent \textbf{RQ4: How often do researchers use statistical results (such as statistical significance) to analyze practical significance?}

During our manual review, we tried to identify papers reporting practical significance, however, practical significance is a complex construct to operationalize. Consequently, we did not reach agreement on how to systematically extract practical significance, hence did not include it in our automated extraction.

In retrospective that is reasonable since recent studies raise the issue that ESE researchers lack a systematic approach to report practical significance~\cite{Krishna2018smells_analytics,Tantithamthavorn2018icse-seip,Torkar2018utility_model}. ESE guidelines suggest using effect size measures or coefficient analysis of a multiple linear regression model to support claims on effect size and, hence, convey such effect as the practical impact of an investigated approach.

However, such practices alone do not capture context information that is relevant for technology transfer of techniques. Therefore, researchers should consider the context related to adoption or application of techniques, etc. 

In order to foster practical significance in ESE, we proposed a conceptual model for a statistical analysis workflow. This model shows how the different statistical methods should be connected and offers guidelines to become aware and overcome common pitfalls in statistical analysis. Moreover, the statistical inference from these methods should lead to practical significance upon input provided by the practitioner's context. Our model also fosters reproducible research by reporting on current reproducibility guidelines from ESE literature. The outcome of the presented practices should feed back into the workflow as empirical data for future studies.

\subsection{Threats to validity}

\noindent \textbf{Conclusion validity:} Our conclusions and arguments rely, predominantly, on two elements: $i$) Krippendorff's $\alpha$, and $ii$) descriptive statistics, visualization and trends in our extracted data. Threats for the keyword extraction also affect a series of validity threats regarding the semi-automatic review. Therefore, to reduce threats concerning our keyword extraction, we systematically argued about the suitability of Krippendorff's $\alpha$ to calculate our inter-rater reliability, as opposed to alternative statistics, e.g., Cohen's $\kappa$, Cohen's $d$, and Cronbach's $\alpha_C$. Furthermore, we were conservative in establishing agreement only on subcategories of the questionnaire with an explicit match from two reviewers, allowing for the exclusion of unreliable keywords for the semi-automatic extraction.

For the descriptive statistics one of the threats is having misleading visualization due to samples, from each journal, varying significantly in the number of papers published per year, e.g., JSS and TOSEM with, respectively, 130 and 15 papers in 2006. For visualization purposes, we used stacked bar charts to visualize the journals within each category, but we avoided clear comparison between journals because each venue involves specific special issues, topics of interests, etc. Moreover, we include absolute numbers for papers identified for the main categories for each year. Ultimately, we focused our discussion on the trends of the papers altogether (see Section~\ref{sec:results}), since focusing on specific statistical method extraction or year can be significantly affected by, respectively, false negatives and false positives.\\

\noindent \textbf{Construct validity:} Our evaluation of the proposed conceptual model relied on the sample of journals used. Even though they are a subset of existing peer-reviewed publication venues, they comprise five popular and top-ranked SE journals. Also, the same sample has been used in similar ESE studies~\citep{Kitchenham2019practices}. An alternative would have been to include conference papers as well, but we argue that the limited number of pages available to papers in conferences could hinder researchers to report thoroughly on their empirical studies, e.g., not including enough details on the choice and usage of statistics. Nonetheless, we intend to extend these constructs in future work.

Similarly, our keywords are also a subset of the existing options for statistical methods, and one can argue whether they are representative. Besides, e.g., tests commonly used in empirical software engineering literature, we extended our keyword extraction to include tests also used in other disciplines, such as psychology. Therefore we cater for a thorough set of tests with corresponding categories (parametric, nonparametric, etc.) that can be used in software engineering research. Some analyzers can be refined, such as the effect size analyzer that can be improved to detect discussions around coefficients of a regression model.

However, including more keywords, at this stage, would represent a liability to the manual extraction, since that would increase the effort of the manual extraction and, in turn, the input for the semi-automatic extraction. With our current choice of keywords, we tried to strike a balance between the diversity of the elements extracted and a feasible analysis for each (and across) category. We further clarified this choice in Section~\ref{sec:semi-auto}.

Finally, we do not claim that our conceptual model is complete concerning steps in a statistical analysis workflow. We anticipate that it is inclusive of the main elements of a statistical analysis and it can be adapted over time when more evidence can be assembled. We do, however, claim that the different aspects are relevant (have been used over the years in ESE research) and that they connect with each other (i.e., one contributes\slash hinders the other).\\

\noindent \textbf{Internal validity:} Our discussion on internal validity threat comprises, primarily, the collection of data using APIs from publishers, and the validation of the extraction tool. Additionally, one can argue that the manual extraction is subjective to internal validity threats, and we agree with that statement. However, most of the threats to validity involved in the manual extraction have been mitigated by the rigorous extraction process and discussion between reviewers, where the outcome is the $\alpha_K$ statistic for the different categories (detailed in Section~\ref{sec:manual}). Therefore, we focus our internal validity discussion on the instruments used for data collection and extraction.

In order to handle data collection, we used REST APIs provided by the corresponding publishers (Springer, IEEE Xplore, and Elsevier)\footnote{See \url{https://dev.springer.com}, \url{http://ieeexplore.ieee.org/gateway/}, and \url{https://dev.elsevier.com/api_docs.html}, respectively.} to collect meta-data about the papers or the articles themselves.\footnote{Only some of the APIs allowed downloading the articles by requests. For instance, no API was found for the ACM Digital Library. Therefore, some of the papers had to be downloaded manually by the authors.} To mitigate the risk of having a mixed approach for data collection, we checked that all downloaded papers were not corrupted by $i$) sampling and verifying the contents of files and, $ii$) automatically checking both the size of the file (in kilobytes) and the article's meta-data (number of words, year, issue, etc.) Files with corrupted content or mismatching information were downloaded again manually. To validate that the summary information captured by the tool had a reasonable accuracy we validated some tags manually, as described in detail in Section \ref{subsec:tool:validation}.\\

\noindent \textbf{External validity:} There are two aspects to evaluate regarding our study's external validity, both pertaining the extent to which our sample enables generalization. The first aspect is related to our sample size (5,196 papers) and the time frame (2001--2015), such that both allowed us generalization regarding the trends seen in those journals. Besides increasing the effort in data collection (i.e., access to the published papers), we believe that increasing the number of years to include papers before 2001 would not add significantly to our conclusions, since we would consequently increase threats, i.e., internal validity, concerning the extraction tool. 

The second aspect of external validity is to broaden the sampling and include more publication venues such as other journals, and eventually proceedings from conferences. We decided to limit our sample to five journals to accommodate variations in the data extraction (manual and semi-automatic). For instance, by adding more journals (e.g., SOSYM,\footnote{International Journal on Software and Systems Modeling} SQJ,\footnote{Software Quality Journal} etc.), we would have to control the higher gap between topics of interest between journals, which could eventually lead to trends unrelated to the use of empirical methods, e.g., rather due to the technical specifics of the papers.

Similarly, including conferences would yield two inconsistent constructs (journals and conference papers) that, even though acting on the same object (i.e., a research paper), could affect the tool's performance concerning detecting false positives at this stage of implementation. In summary, we argue that the sample size and choice of journals enables generalization of our findings, even though future research should expand towards different areas within software engineering by including conference proceedings.

\section{Conclusions and future work}\label{sec:concl}
This paper analyses what are the main statistical methods used across 15 years of empirical software engineering research. Our analysis includes data from five well-known and top-ranked software engineering journals (TSE, TOSEM, EMSE, JSS and IST) between 2001--2015. We performed our review in two distinct stages where we, first, performed a manual review on all papers published in the selected journals for the year of 2015 and then, used the outcome of this manual review to design and use a semi-automatic tool to extract data on a larger pool of papers from the same venues.

Our results reveals that, overall, the use of statistical analysis of quantitative data is on the rise with an especially steady increase in the last years. More specifically in the last five years, there has been an increased use of nonparametric statistics as well as of effect size measurements. Conversely, we identified certain problems such as the amount of parametric tests identified that do not explicitly report any test for normality.

We used the extracted data to create a conceptual model for a statistical analysis workflow that ESE researchers can use to guide their choice and usage of statistical analysis. This model includes a set of guidelines to raise awareness of the main pitfalls in different stages of a statistical analysis (e.g., performing power analysis, or reporting on effect sizes) and support researchers in achieving and reporting practical significance of their results. For the latter we emphasize the need to consider one or more practical context where the investigated technique should be applicable. Consequently, researchers become able to cater to practitioners' expectations when adopting, e.g., techniques.

We also provide suggestions of different techniques that can complement the current statistical toolkit of ESE researchers such as using Bayesian analysis able to provide more informative results, and imputation techniques to overcome missing data points.

There are two main areas of future work: $i$) refining the semi-automated extraction, and $ii$) expanding our analysis to include more venues or analysis techniques. The first is ongoing work, while our tool, named \texttt{sept}, is updated with refinements on the existing set of analyzers in order to improve its keyword detection capabilities, as well as including more extractors able to support humans in mechanic and tedious reviewing tasks such as submitting double-blind papers or including acknowledgements to funding agencies and projects. A natural addition would also be to use more advanced Natural Language Processing to improve automated extraction. The second area of future work involves expanding our pool of papers to include more samples from both other journals and conferences in SE, as well as aggregating results based on different fields and investigating how statistical methods are used in different types of empirical studies.

\section{References}
\bibliography{main}
\end{document}